\documentclass[preprint,12pt]{elsarticle}  

\usepackage{amssymb} 
\usepackage{amsmath}
\usepackage[dvipsnames]{xcolor} 
\usepackage{wrapfig}
     
\usepackage{ulem} 

\newcommand{\uinc}{u^{\mathrm{in}}} 
\newcommand{\uincbg}{u^{\mathrm{in}}} 
\newcommand{\uscbg}{u^{\mathrm{sc}}_{\mathrm{bg}}} 
\newcommand{\rmi}{\mathrm{i}} 
\usepackage{hyperref}

\usepackage{caption} 

\usepackage[labelformat=simple]{subcaption} 

\begin{document}


\begin{frontmatter}



\title{Solving forward and inverse wave scattering via boundary integral equations and deep learning. Applications to cloaking design.}

\author[insa]{Camille Carvalho}
\author[ucm]{Elsie Cortes\corref{cor1}\fnref{fn1}}
\author[ucm]{Symeon Papadimitropoulos}
\author[ucm]{Chrysoula Tsogka}

\cortext[cor1]{Corresponding author}
\fntext[fn1]{Email: ecortes7@ucmerced.edu}


\affiliation[ucm]{
  organization={University of California, Merced},
  addressline={5200 Lake Rd},
  city={Merced},
  state={CA},
  postcode={95343},
  country={USA}
}

\affiliation[insa]{
  organization={INSA Lyon, Centrale Lyon, Université Lyon 1, Université Jean Monnet, CNRS, ICJ UMR5208},
  city={Villeurbanne},
  postcode = {69621},
  country = {France}
}


\begin{abstract}

We propose a deep learning framework based on an encoder-decoder architecture for the design and evaluation of cloaking devices, demonstrated in this work for two-dimensional wave propagation governed by the Helmholtz equation.  The cloaks under consideration are concentric layered media surrounding the object, whose geometry and material parameters determine the scattering response.  We consider circular and object-fitted layer configurations and parameterize all designs by the layer thicknesses, enabling a unified representation for direct comparison of different cloaks for the same object.  Training data are generated using a boundary element formulation suitable for geometries where analytic solutions are not available, and neural networks are trained with standard hyperparameters on geometry-specific datasets.  The proposed approach is applied to circular, star-shaped, and kite-shaped objects.  Results show that object-fitted configurations consistently outperform simpler circular-layer designs in scattering reduction, highlighting the importance of geometry in cloaking performance.  Overall, we present a flexible, data-driven approach for systematic comparison of cloaking strategies, with potential extension to more complex geometries and broadband settings.

\end{abstract}

%


\end{frontmatter}



\section{Introduction}
\label{sec1} 

Invisibility has long captured the imagination of both storytellers and scientists alike.  Though discussions of it are often confined to fiction, 
cloaking technologies have attracted growing scientific attention due to the diverse range of applications beyond the pursuit of optical invisibility 
\cite{Lee2021}.  In addition to rendering objects undetectable to visible light, cloaking models have been explored for protecting structures from 
seismic vibrations, concealing sensors to enable noninvasive medical imaging, and controlling acoustic or thermal wave propagation \cite{cassier2022activeexteriorcloaking2d}.  The fundamental 
concept across these applications involves the reduction of wave scattering with materials so the field exterior to the object appears unimpeded.  
For instance, this manipulation of the wave field can be done via transformation optics, an approach to control field behavior by prescribing the spatial variation of material parameters according to a coordinate transformation \cite{pendry2006}.  
The approach we are more interested in utilizing consists of using the so-called metamaterials: constructions of thinly layered contrasting materials that, when assembled, exhibit particular effective optical properties conducive to cloaking \cite{alu2005,cai2007}. 
The optimal metamaterial design for a cloak depends strongly on the geometry and material composition of the object to cloak, and identifying the appropriate set of materials in each context is difficult. 
Determining the cloak design parameters in each case typically involves solving individual complex scattering problems, which is computationally expensive and time-consuming.

In this work, we specifically consider electromagnetic wave cloaking of time-harmonic waves in 2D, which consists of a multilayered system representing a construction of multiple cloaking shells.
For object and cloak configurations with simple shapes, like circles and ellipses, analytic solutions are available.  
However, more complex object-cloak configurations require alternative approaches like finite element and finite difference methods, which can accommodate a wide range of geometries \cite{schweigerFEM1997,VolakisFEM,frehnerComparisonFiniteDifference2008}.
These volumetric methods, however, require meshing the entire computational domain and truncating it with absorbing boundaries.  
This makes them less suitable for applications where we are only interested in the far-field wave behavior, such as cloaking.
 Alternatively, here we consider boundary element methods (BEM), which only require 
discretization of the object and cloak interfaces \cite{Bonnet1999-bf}.
BEM are particularly advantageous because solutions in the volume can be evaluated at arbitrary points without redefining the underlying mesh of the interfaces \cite{liuBEMAcousticWave2019}.
This reduction of the problem's dimensionality is useful not only for our 2D setting, but significantly efficient for extensions to 3D as well. 
However, BEM introduce their own numerical challenges, including weakly and nearly singular kernels in the boundary integral formulation that require specialized treatment; this is the so-called close evaluation problem \cite{barnett_evaluation_2014}. 
These become especially pronounced when modeling metamaterial cloaks with thin, closely spaced layers. Indeed, in that case, near-field interactions become essential to capture, and failing to do so significantly impacts accuracy.  
We address these limitations in this work, presenting an approach to handle this close evaluation problem to enable efficient and accurate approximation of the solutions.

BEM is only employed in this work, however, to solve the forward scattering problem.  
For metamaterial design, it is important to adopt a method which is better-suited for dealing with the complexities of solving the inverse scattering problem, which is highly nonlinear and subject to non-uniqueness of solutions.   
There are a number of traditional inverse scattering solvers that are employed depending on the problem and, in some cases, the problem's constraints.  
For example, PDE solvers such as nonlinear least squares are general, flexible, and applicable to high-contrast scattering problems, but they often require repeated expensive forward solves, stabilizers, and preconditioners which make them implementation-intensive \cite{colton2015inverse}. 
This has motivated many recent works in neural network approaches for the inverse scattering problem, in which models learn direct relationships between wave field data to medium parameters \cite{liuGenerativeModelInverse2018,maDeepLearningEnabledOnDemandDesign2018,maProbabilisticRepresentationInverse2019}. 
Notably, data-driven approaches have been shown to perform well in high-contrast regimes specific for modeling metamaterials. By leveraging the stability and well-posed nature of the forward problem to generate training datasets, they enable the learning of more robust and efficient mappings for the inverse problem \cite{metamaterialdesignMachineIntelligence,zhangDataDrivenApproaches2025}.

Among the various deep learning models, the architecture of the encoder-decoder framework has been previously shown to be particularly effective for acoustic and electromagnetic cloaks \cite{liuTrainingDeepNeural2018,Kiarashinejad2020,Ahmed2021}.
We adopt this structure due to its demonstrated success and its structural simplicity, and explore the adaptation of the approach to a wide range of object-cloak configurations. 
We emphasize the use of a general architecture with minimal hyperparameter tuning, enabling the same framework to be applied across multiple cloaking scenarios.
This enables direct comparison of different configurations for a chosen object to cloak, which is ideal for identifying the most optimal cloaking device designs.
{With the proposed BEM approach combined with the encoder-decoder architecture, in this paper we provide a full and flexible pipeline to solve forward and inverse wave scattering, that could be applied to other applications than cloaking.}

The work presented here is organized as follows.  In Section~\ref{sec2}, we introduce the formulation of the multilayer scattering problem, outlining the associated parameters and governing equations. In Section \ref{sec3} we present the neural network framework employed, including details on the hyperparameters chosen for our models. Section~\ref{sec4} describes the methodology for constructing datasets used to train and test our models.  In Section~\ref{sec5}, we demonstrate the application of the network to various object cloaking problems, evaluating the ability of the constructed neural network to output optimal cloaking designs.  We conclude in Section~\ref{sec6} with a summary of the design implications emerging from the results and an outline of directions for future work. 

\section{The multilayered scattering forward problem}
\label{sec2}

\begin{figure}[!h]
  \centering
  \includegraphics[width=0.65\textwidth]{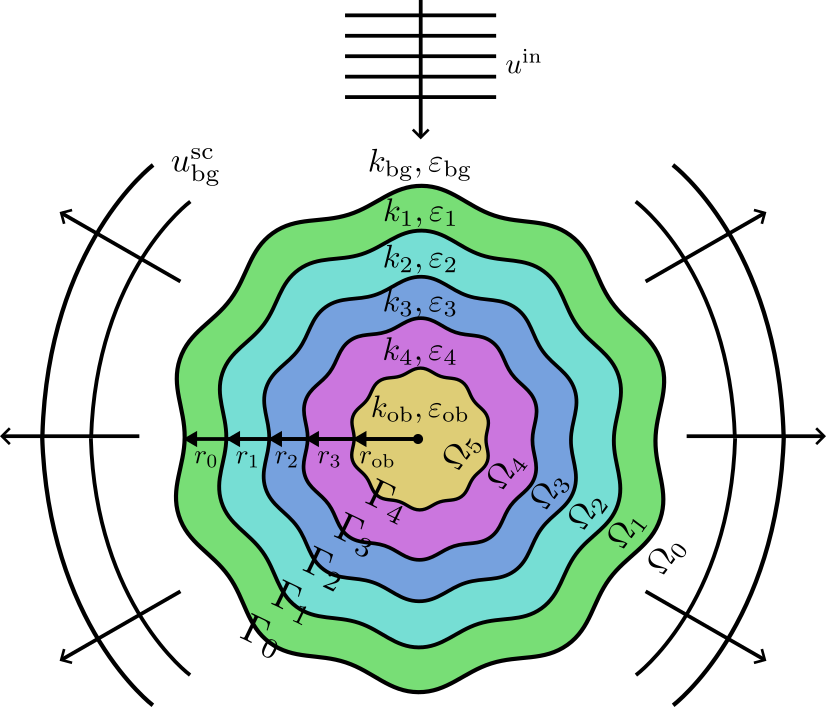}
\caption{Example of a 5-layer configuration with \textit{star-shaped} layers.
The wave numbers \(k_j\), permittivities \(\varepsilon_j\), and radii \(r_j\) are labeled, including an innermost object defined by \(r_{\mathrm{ob}}\), \(k_{\mathrm{ob}}\), and \(\varepsilon_{\mathrm{ob}}\) in region \(\Omega_5 = \Omega_{\mathrm{ob}}\).
An incident plane wave \(\uincbg\) generates a scattered field \(\uscbg\) in the background medium in \(\Omega_0 = \Omega_{\mathrm{bg}}\), defined by \(k_{\mathrm{bg}}\) and \(\varepsilon_{\mathrm{bg}}\).}
  \label{fig:problem structure}
\end{figure}

We consider wave propagation through a configuration of \(N + 1\) concentric, material regions \(\Omega_j\) separated by smooth, nested interfaces \(\Gamma_j\) for \(j = 0, \hdots, N\).
Each interface \(\Gamma_j\) is assumed to admit a {global polar parameterization in terms of \((r, \theta)\), and depends on a characteristic radius $r_j$. For example, we consider the parameterization $(x(\theta), y(\theta)) = (r_j \cos(\theta),r_j \sin (\theta))$ when circular, but more generally we will consider interfaces that can be parameterized as $(x(\theta), y(\theta)) = (x(r_j, \theta) \cos(\theta),y(r_j,\theta) \sin (\theta))$. For the cloak, we assume the interfaces admit the same type of parameterization. In other words, we select one shape and add concentric layers of the same shape. The radii are} ordered as \(r_0 > r_1 > \hdots > r_{N-1}\), ensuring that the interfaces are closed, concentric, and non-intersecting.
Each region is characterized by permittivity \(\varepsilon_j >0\).  We assume unit magnetic permeability across all regions, \(\mu_j = 1\) for \(j = 0, \hdots, N\), so that the material properties of each region are fully described by the permittivity.
The wavenumber in the \(j\)-th region is given by \(k_j = \omega \sqrt{\varepsilon_j}\), where \(\omega\) denotes the fixed frequency of the time-harmonic regime.  The exterior region \(j = 0\) represents the background medium, defined by wavenumber \(k_0 := k_{\rm{bg}}\) and permittivity \(\varepsilon_0 := \varepsilon_{\rm{bg}}\).
The innermost region \(j = N\) corresponds to the object to be cloaked, characterized by radius \(r_{N-1} := r_{\rm{ob}}\), wavenumber \(k_N :=k_{\rm{ob}}\), and permittivity \(\varepsilon_N:=\varepsilon_{\mathrm{ob}}\).  An example of a configuration for a $N=5$ layer problem (or a 4-layer cloak with one object layer) is shown in Figure \ref{fig:problem structure}.

While considering electromagnetic wave scattering under Transverse Magnetic polarization, Maxwell's equations educe to a Helmholtz problem for the magnetic field \(u\). Assuming time-harmonic dependence of the form
\[u_j(\mathbf{x},t) = u_j(\mathbf{x})\exp({\rmi\omega t})\]
in each region $\Omega_j$, and henceforth denoting the spatial component \(u_j(\mathbf{x})\) simply by \(u_j\), we obtain the following multi-layer transmission problem:
\begin{equation}\label{eq:fwd_pb}
\begin{aligned}
& \text{Helmholtz equation: }\\
  &   \nabla^2 u_j + k^2_j u_j = 0 \quad \text{ in } \Omega_j, \quad j = 0, \dots N\\
  & \text{Transmission conditions: }\\
& \left[ u\right]_{\Gamma_j} = 0, \text{ and } \quad \left[ \frac{1}{\varepsilon}\frac{\partial u}{\partial n_j}\right]_{\Gamma_j} = 0, \quad \text{ on } \Gamma_j, \quad j = 0, \dots N-1\\
& \text{Sommerfeld radiation condition: }\\
& \lim_{r \to \infty} \sqrt{r} \left( \frac{\partial u_{\rm{bg}}^{\rm{sc}}}{\partial r} - \rmi k_{\rm{bg}}u_{\rm{bg}}^{\rm{sc}} \right) = 0.
  \end{aligned}
\end{equation}
where $\left[ f\right]_{\Gamma_j}=f_{j+1} - f_j$ denotes the jump of  a function\(f\) across the interface $\Gamma_j$ and \(n_j\) denotes the unit outward normal on \(\Gamma_j\). \\
Let us make some remarks:
\begin{itemize}
    \item We chose to consider a penetrable object. We may also employ other boundary conditions, such as Neumann or Dirichlet, on \(\Gamma_{N-1}\) if the object is non-penetrable.
    \item We chose the incident field \(\uinc\) to be a plane wave of the form \(\rm\exp{(\rmi k_{\rm{bg}} \mathbf{a} \cdot \mathbf{x})}\) for \(\mathbf{a} := (\cos\alpha, \sin\alpha)\).  In this work, we fix the incident angle \(\alpha = \pi/2\) so that the plane wave propagates in the negative \(y\)-direction, though all results hold for any choice of \(\alpha\).
\end{itemize}
For the forward problem, the goal is to characterize the wave propagation and scattering behavior of the solutions \(u_j(\mathbf{x})\) by this layered structure for a chosen object and background medium in order to inform the design of effective cloaking configurations.  We will utilize BEM to solve the forward problem, but the focus of this work is developing an approach to solve the inverse problem for various layer geometries.

\section{The deep learning approach for multilayered scattering}
\label{sec3}
\subsection{Problem setup}
\begin{figure}[!b]
    \centering
    \includegraphics[width=0.75\linewidth]{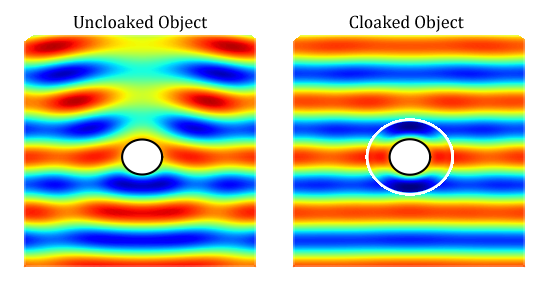}
    \caption{The real part of the wave field solution is shown. (Left) Incoming wave field scatters upon hitting the object. (Right) The object is cloaked by a surrounding metamaterial.}
    \label{fig:cloaking_example}
\end{figure}
We seek to solve the multilayered scattering inverse problem: given the properties of the background medium and the object, recover the radii and permittivities of the cloaking layers defined by regions \(\Omega_1, \hdots, \Omega_{N-1}\). 
A successful cloak renders the object undetectable, i.e., the wave field in the background medium is undisturbed with little to no scattering, so that the total field is approximately equivalent to the incident field.  That is, $u_{\rm{bg}}(\mathbf{x}) \approx \uincbg(\mathbf{x})$ (or the scattered field \(\uscbg(\mathbf{x}) \approx 0\)), as illustrated in Figure~\ref{fig:cloaking_example}. 
We assume a priori knowledge of the background permittivity \(\varepsilon_{\rm{bg}}\) and the number of layers \(N\) so that the inverse problem is posed on a consistent parameter space.  
This allows us to focus on recovering the geometric and material parameters of the cloaking layers within a unified representation that facilitates comparison across different configurations.

Even with these constraints, however, solving the inverse scattering problem is challenging because it is high-dimensional, ill-posed, and nonlinear  \cite{ra2001problems}.  
The wave field for the multilayer problem is defined over a spatial domain, and the discretization of this domain leads to a high-dimensional parameter space.
The problem is ill-posed because the mapping between wave field data to material parameters is typically non-unique and unstable with respect to noise, meaning small measurement errors may lead to large variations in the reconstructed solutions.
It is also nonlinear because the wave field solution is obtained by solving a PDE whose coefficients depend on unknown layer parameters, with coupling across interfaces through transmission conditions.
To  address these challenges we employ a deep-learning approach, in particular, an encoder-decoder neural network that learns directly from data. 
It learns how to capture the nonlinear relationship between material parameters and wave field solutions without relying on an explicit analytic model.

In this architecture, the decoder approximates the forward problem, mapping cloaking layer parameters to the resulting scattered wave field. 
The encoder solves the inverse problem, compressing a given wave field into a compact latent representation from which the layer parameters are reconstructed. 
Rather than training the encoder in isolation, we train it through the combined network, using the pre-trained decoder to assess whether the predicted parameters faithfully reproduce the input field. 
This setup effectively mitigates the ill-posedness of the inverse problem, as the encoder is guided towards physically consistent parameter predictions.

Encoder-decoder architectures of this type have shown significant promise for layered geometries in prior work \cite{Ahmed2021,Wu2022}. 
We would like to extend the use of this architecture to explore cloaking devices for a wider range of different object and cloak geometries than previously examined. Indeed, it is common to consider circular or elliptical layers as an analytic solution is available for those cases. However, the use of such approaches for other types of geometries have not been explored, to our knowledge.
The following subsections describe the key network components and hyperparameter choices in detail.

\subsection{Decoder model for the forward problem}
Since all interfaces will be parameterized the same way, the relevant parameters for the layered structure are the layer thicknesses \(\Delta\mathbf{r} = (r_j - r_{j+1})^{N-2}_{j=0}\), and permittivities \(\mathbf{e} =(\varepsilon_j)^{N-1}_{j=1}\) respectively. We first train a decoder network to approximate the forward problem, mapping the physical parameters of the multilayer structure to the resulting scattered wave field evaluated at a fixed set of points in the background medium. Working with the layer thicknesses $\Delta\mathbf{r}$ rather than the radii $r_j$ directly is advantageous because the thicknesses are naturally positive quantities, whereas using the radii would require explicitly enforcing the ordering constraints $r_0 > r_1 > \cdots > r_{N-1}$ during training, which is considerably more difficult. 

\begin{figure}[!h]
    \centering
    \includegraphics[width=0.55\textwidth]{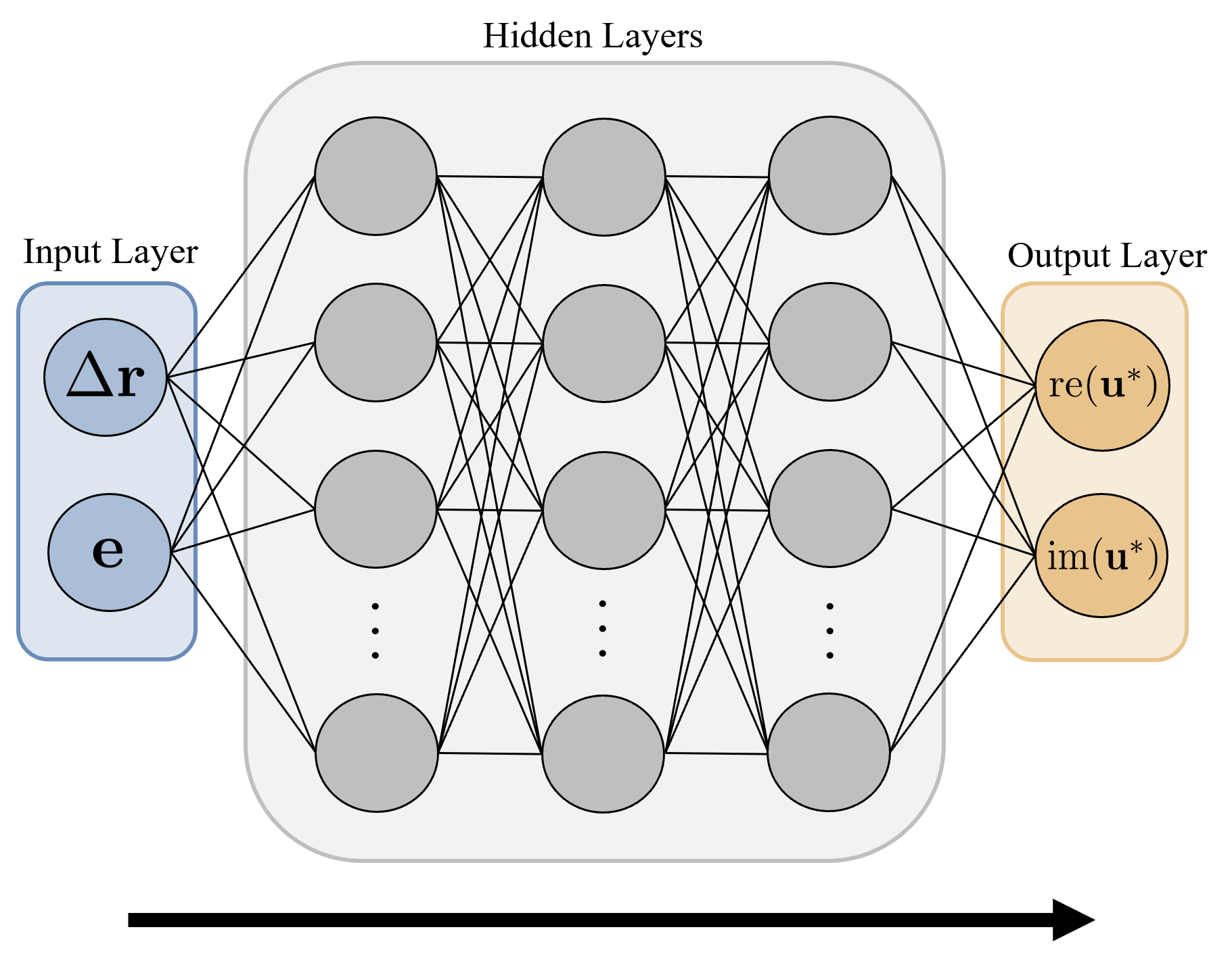}
    \caption{Decoder architecture mapping layer thicknesses \(\Delta\mathbf{r} = (r_j - r_{j+1})^{N-2}_{j=0}\) and permittivities \(\mathbf{e} =(\varepsilon_j)^{N-1}_{j=1}\)  to the real and imaginary parts of the wave field \(\mathbf{u^*}\) at a set of discrete spatial points.}\label{fig:decoder model}
\end{figure}

The general structure of the decoder is shown in Figure~\ref{fig:decoder model}. The wave field is separated into its real and imaginary components $\text{re}(\mathbf{u})$ and $\text{im}(\mathbf{u})$, resulting in an output layer that is twice the size of the point distribution in the background medium. Because the forward problem has a unique solution for each set of parameters, training a model to approximate it is straightforward and typically converges well.

\begin{table}[h]
\centering
\begin{tabular}{|l | l | }
\hline
Parameter & Value \\
\hline
Batch size & 128 \\
Hidden layer sizes & [16, 32, 64] \\
Layer type & Fully connected \\
Activation function & GELU \\ 
Optimizer & Adam \\
& (learning rate =0.001, weight decay=\(1\mathrm{e}{-5}\))\\
\hline
\end{tabular}
\caption{Hyperparameter configuration of the neural network for the forward problem.}
\label{tab:forward params}
\end{table}

 The hyperparameter configuration of the feedforward neural network we implement for the decoder is summarized in Table \ref{tab:forward params}.
 We use a batch size of 128, meaning the model is updated after processing 128 training samples at a time.  It is a middle-ground batch size that balances computational efficiency and stability of training.
 The selected learning rate and weight decay are also standard values for the Adam optimizer.   
 While these parameters could be tuned for individual problems, we find that a single set of standard values performs well across all cases considered in this work. 
 This is also more consistent with our goal of developing a general architecture applicable to multiple problem settings.

The decoder consists of a fully connected architecture which adopts a low-to-high input to output mapping to support the reconstruction of the high-dimensional wave field from our compact set of parameters.  It is composed of Gaussian Error Linear Unit (GELU) activation functions in the hidden layers.  Unlike ReLU, which is piecewise linear and can suffer from dead neurons, or Tanh, which saturates for large input values, GELU provides smooth, nonlinear gradient responses that improve the network's ability to reconstruct field solutions while maintaining numerical stability during training \cite{lee2023}.  Other standard activation functions were tested, including Sigmoid and LeakyReLU functions, but the GELU activation function showed the best convergence behavior in all test cases.  The resulting decoder is compact yet effective, and forms the backbone of the combined encoder-decoder framework described next.

\subsection{Combined encoder-decoder model for the inverse problem}

\begin{figure}[!h]
\centering
\includegraphics[width=0.9\textwidth]{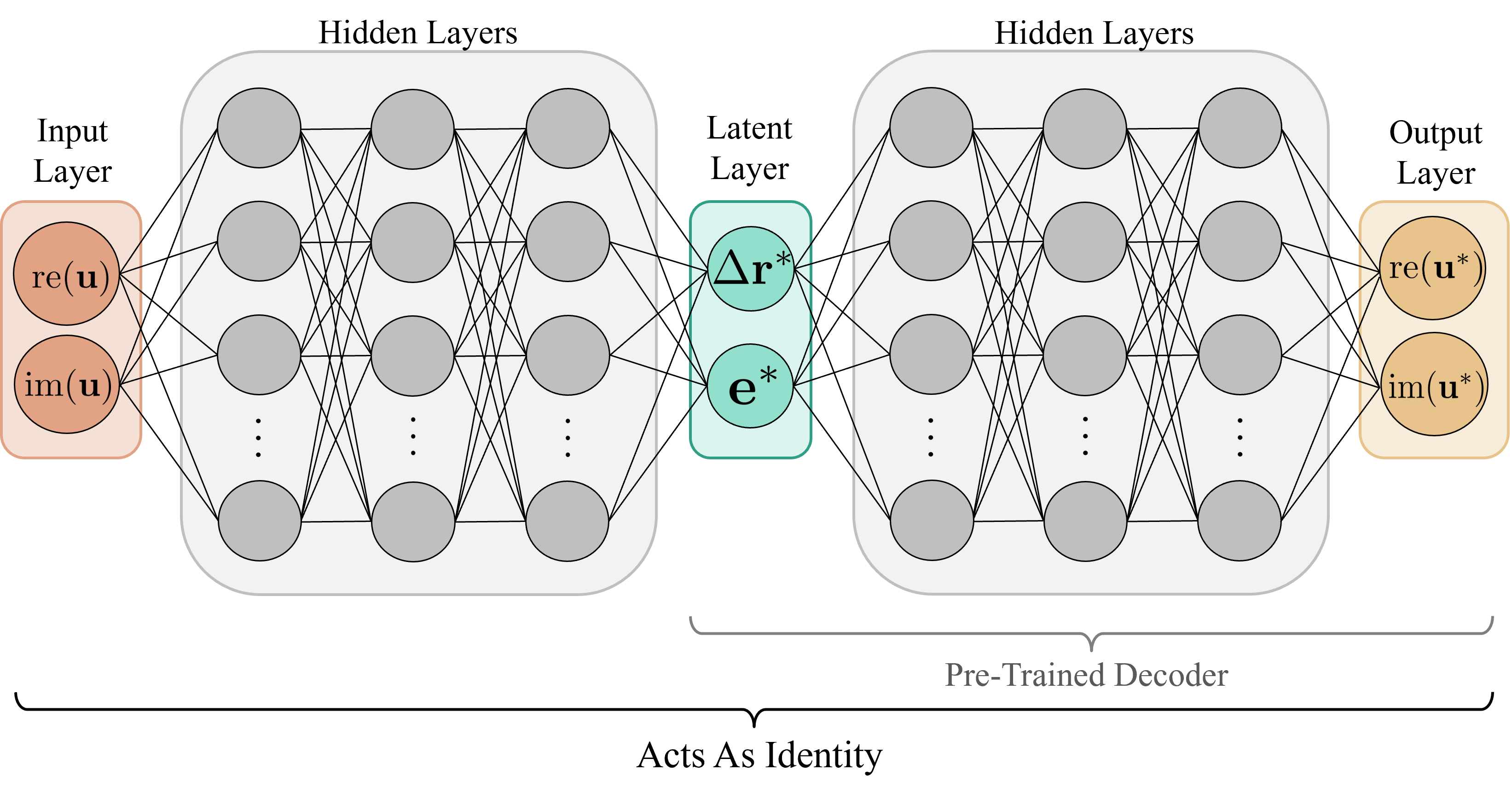}
\caption{Combined encoder-decoder architecture trained to act as the identity. The encoder maps the real and imaginary parts of the wave field \(\mathbf{u}\) to estimates of layer thicknesses \(\Delta \mathbf{r}\) and permittivities \(\mathbf{e}\) while the  pretrained decoder given the layer thicknesses \(\Delta \mathbf{r}\) and permittivities \(\mathbf{e}\) provides an estimate of the real and imaginary parts of the wave field \(\mathbf{u^*}\).}\label{fig:encoder model}
\end{figure}

The encoder solves the inverse problem, mapping a given wave field back to the physical parameters of the cloaking device. As noted earlier, this mapping is not unique: multiple parameter configurations can produce the same, or similar, scattered fields. To handle this ill-posedness, the encoder is not trained in isolation but through the combined encoder-decoder network (see Figure~\ref{fig:encoder model}), where the decoder output is compared to the original input field. This effectively trains the encoder to produce parameters that, when passed through the pre-trained decoder, reproduce the input field as closely as possible.

In an encoder-decoder architecture, the encoder typically uses a symmetric but inverted architecture relative to its partner decoder.  So, in this work, we use fully connected hidden layers of sizes $[64, 32, 16]$. Apart from the activation functions, the remaining hyperparameters follow the same configuration as the decoder (as shown in Table~\ref{tab:inverse params}). For the encoder, we found after rigorous testing that Exponential Linear Unit (ELU) activation functions ensured more robust learning of the inverse mapping than other activation functions, including GELU. This is likely because ELUs avoid the dead neuron problem associated with ReLU, maintain nonlinear responses throughout training \cite{song2024}. ELU generally show faster convergence on larger datasets and were tested for the decoder as well \cite{dubey2022}, but only outperformed GELU when used to train the encoder. Specifically, we construct a vertically-shifted variant of ELU with $\alpha = 1$:
\begin{equation}
   f(x) = \begin{cases} \begin{aligned}
        \alpha x + 1, & \quad \quad  x\geq0\\
(\alpha (\exp^x-1)) + 1, & \quad \quad  x<0, 
    \end{aligned} 
\end{cases}
\end{equation}
which produced the best convergence and output behavior across all problem settings considered, including objects with exotic shapes. In some cases, the EReLU activation function \cite{maurya2023} also produced effective cloaks, though generally not as effective as the shifted ELU.  In the output layer, the softplus function (a smooth approximation of ReLU) is also used to enforce positivity of the predicted parameters consistent with the physical constraints of the problem. 

\begin{table}[t]
\centering
\begin{tabular}{|l | l | }
\hline
Parameter & Value \\
\hline
Batch size & 128 \\
Hidden layer sizes & [64, 32, 16] \\
Layer type & Fully connected \\
Activation function & Shifted ELU (Softplus in output layer)\\ 
Optimizer & Adam \\
& (learning rate =0.001, weight decay=\(1\mathrm{e}{-5}\))\\
\hline
\end{tabular}
\caption{Hyperparameter configuration of the neural network for inverse problem.}
\label{tab:inverse params}
\end{table}

With the encoder-decoder architecture of our network established, we now turn to the next component of the pipeline: data generation.

\section{BEM formulation for data generation}
\label{sec4}
The difficulty of the cloaking task depends on how close to the object the scattering effect must be unobservable. However, as long as little to no scattering is observed beyond a certain distance, the Sommerfeld radiation condition guarantees that the scattered field continues to decay further from the source. It therefore suffices to train on wave field data sampled exclusively in the background medium, outside the subdomains containing the cloaked object and cloaking layers.

To derive the wave field solutions in the background medium, we employ a boundary integral equation (BIE) formulation. 

\subsection{Boundary integral equations for multilayered scattering problem}
In each subdomain $\Omega_j$, the solution to the Helmholtz equation can be represented via a combination of single- and double-layer integral operators $\mathcal{S}_{i,j}$ and $\mathcal{D}_{i,j}$ \cite{kress_boundary_1991}, defined as

\begin{align*}
    \mathcal{S}_{i,j}[\mu](\mathbf{x}) &= \int_{\Gamma_i} \Phi_j(\mathbf{x},\mathbf{y}) \mu(\mathbf{y}) d\Gamma_{i,\mathbf{y}} , &\quad \mathbf{x} \in \Omega_j,\\
        \mathcal{D}_{i,j}[\mu](\mathbf{x}) &= \int_{\Gamma_i} \frac{\partial \Phi_j(\mathbf{x}, \mathbf{y})}{\partial n_i(\mathbf{y})} \mu(\mathbf{y}) d\Gamma_{i,\mathbf{y}} ,&\quad \mathbf{x} \in \Omega_j.
\end{align*}

In these integral operators, \(\Gamma_i\) (for \(i = j - 1\)  or \(j\)) denotes the interface on which they are defined. \(\Phi_j\) is the fundamental solution to the Helmholtz equation in \(\Omega_j\), which in 2D is given by  
\begin{equation}
    \Phi_j(\mathbf{x},\mathbf{y}) = \frac{\rmi}{4}H_0^{(1)}(k_j|\mathbf{x}-\mathbf{y}|)
    \label{eq:fundamental solution}
\end{equation}

where \(H_0^{(1)}\) is the Hankel function of the first kind.
The unknown quantities $\mu$ in \(\mathcal{D}_{i,j}\) and \(\mathcal{S}_{i,j}\) are called density functions.
They define the \textit{boundary data}: we choose to use the standard representation formula, therefore in the double-layer operator, $\mu$ will be the trace $u_j$, and in the single-layer operator it will be the normal trace \(\tfrac{\partial u_j}{\partial n_i}\).
For simplicity, we denote the traces \(u_j(\mathbf{y})\) and \(\tfrac{\partial u_j}{\partial n_i}(\mathbf{y})\) in $\mathcal{D}_{i, j}$ and $\mathcal{S}_{i, j}$ respectively by
\begin{equation*}
u_{i, j} := u_j(\mathbf{y}), \quad \text{and}\quad  \partial_n u_{i, j} := \frac{\partial u_j}{\partial n_i}(\mathbf{y}), \quad \quad \mathbf{y} \in \Gamma_i.
\end{equation*}
Using the integral operators, we obtain the following boundary integral representations of the solutions \(u\):
\begin{equation}
\resizebox{0.9\textwidth}{!}{$
\begin{aligned}
    u_0(\mathbf{x}) &= u^{\mathrm{in}}(\mathbf{x}) + \mathcal{D}_{0,0}[u_{0,0}](\mathbf{x}) + \mathcal{S}_{0,0}[\partial_n u_{0,0}](\mathbf{x}) & \text{in } \Omega_0 \\
    u_j(\mathbf{x}) &= -\mathcal{D}_{j-1,j}[u_{j-1,j}](\mathbf{x}) + \mathcal{S}_{j-1,j}[\partial_n u_{j-1,j}](\mathbf{x}) \\ 
    &\quad + \mathcal{D}_{j,j}[u_{j,j}](\mathbf{x}) - \mathcal{S}_{j,j}[\partial_n u_{j,j}](\mathbf{x}) & \text{in } \Omega_j \\
    u_N(\mathbf{x}) &= -\mathcal{D}_{N-1,N}[u_{N-1,N}](\mathbf{x}) + \mathcal{S}_{N-1,N}[\partial_n u_{N-1,N}](\mathbf{x}) & \text{in } \Omega_N \\
\end{aligned}
$}
\label{eq:bie_layers}
\end{equation}

for \(j = 1, \hdots, N - 1\).

The solution in each region \(\Omega_j\) is determined from the traces of \(u\) and normal trace \(\partial_n u\) on each interface. To obtain the unknown traces \(u_{i,j}\) and \(\partial_n u_{i,j}\), we construct a system of equations defined on the interfaces in terms of layer potentials.  These layer potentials are found by taking the limits of the single- and double-layer integral operators on each side of the boundary $\Gamma_j$, and are called \(S_{i,j}\) and \(D_{i,j}\) respectively.  While the single-layer operator is continuous, the double-layer operator exhibits a jump discontinuity: for a point \(\mathbf{x}^b\) on the boundary, it produces an additional \(\pm \frac{1}{2}\mu(\mathbf{x}^b)\) term such that

\begin{equation*}\mathcal{D}_{i,j}[\mu](\mathbf{x}) \underset{\mathbf{x}\rightarrow \mathbf{x}^{b, \pm}}{\longrightarrow}\pm \frac{1}{2}\mu(\mathbf{x}^b) + D_{i,j}[\mu](\mathbf{x}^b).\end{equation*}

Applying the transmission conditions in ~\eqref{eq:fwd_pb} to the boundary integral representation equations \eqref{eq:boundaryrep} yields a system of $2N$ equations and $2N$ unknowns.  Denoting a boundary point \(\mathbf{x}\) on \(\Gamma_j\) as \(\mathbf{x}^b_j\), we make the choice to keep the traces $u_{j,j}$ and normal traces $\partial_n u_{j,j}$, so that the BIE system to solve is:
\begin{equation}
\resizebox{0.93\textwidth}{!}{$
\begin{aligned}
    \tfrac{1}{2}u_0(\mathbf{x}_0^b)& = u_0^{\mathrm{in}}(\mathbf{x}_0^b) + D_{0,0}[u_{0,0}](\mathbf{x}_0^b) + S_{0,0}[\partial_n u_{0,0}](\mathbf{x}_0^b)\\
    \tfrac{1}{2}u_j(\mathbf{x}_{j-1}^b) &= -D_{j-1,j}[u_{j-1,j-1}](\mathbf{x}_{j-1}^b) + \frac{\varepsilon_{j}}{\varepsilon_{j-1}}S_{j-1,j}[\partial_n u_{j-1,j-1}](\mathbf{x}_{j-1}^b) \\
    & \qquad \mathcal{D}_{j,j}[u_{j,j}](\mathbf{x}_{j-1}^b) - \mathcal{S}_{j,j}[\partial_n u_{j,j}](\mathbf{x}_{j-1}^b)\\
    \tfrac{1}{2}u_j(\mathbf{x}_j^b) &= -\mathcal{D}_{j-1,j}[u_{j-1,j-1}](\mathbf{x}_j^b) + \frac{\varepsilon_{j}}{\varepsilon_{j-1}} \mathcal{S}_{j-1,j}[\partial_n u_{j-1,j-1}](\mathbf{x}_j^b) \\
    &\qquad  + D_{j,j}[u_{j,j}](\mathbf{x}_j^b) - S_{j,j}[\partial_n u_{j,j}](\mathbf{x}_j^b)\\
    \tfrac{1}{2}u_N(\mathbf{x}_{N-1}^b) &= -D_{N-1,N}[u_{N-1,N-1}](\mathbf{x}_{N-1}^b)  + \frac{\varepsilon_{N}}{\varepsilon_{N-1}}S_{N-1,N}[\partial_n u_{N-1,N-1}](\mathbf{x}_{N-1}^b)
\end{aligned}
$}
\label{eq:boundaryrep}
\end{equation}

On the boundaries \(\Gamma_j\) for \(j = 1, \hdots, N - 2\), the integral equations incorporate both the operators associated with the other interface and the layer potentials defined on the same interface where the boundary solution is being evaluated.

The size of the resulting system is determined by the number of interfaces $N$ and the number of boundary points $M$ we use to discretize the integrals. Once the unknown traces and normal traces are obtained, the wave field in the background domain may be evaluated using the boundary integral representation for $u_0(\mathbf{x})$. While we may simply need to compute the field in the background (and therefore have knowledge of only $u_{0,0}, \partial_n u_{0,0}$), it is necessary to fully solve \eqref{eq:boundaryrep} due to coupled terms. In order to solve the BIE system, we make use of standard quadrature rules such as the Kress Product Rule \cite{kress_boundary_1991}, yielding spectral accuracy to treat weakly singular integrals. Here, the Hankel function and its derivatives can exhibit a logarithmic singularity as $|\mathbf{x}_j^b - \mathbf{y}| = 0$, which requires specific treatment. We refer to \cite{kress_boundary_1991, CCCT2026} for some details. Once the boundary data is computed, we approximate the wave field solution with the periodic trapezoidal rule (with $M$ quadrature points), which yields spectral convergence in $M$ for smooth, closed boundaries \cite{Bonnet1999-bf}.

For points near but not on the boundary, however, the Hankel function and its derivatives become nearly singular as $|\mathbf{x} - \mathbf{y}| \to 0$. This is known as the \textit{close evaluation problem}. The phenomenon is essential to handle as large errors may occur if not addressed explicitly. Here, it may appear twice: (i) when evaluating the field with \eqref{eq:bie_layers} especially close to a boundary, (ii) when solving the BIE system \eqref{eq:boundaryrep} when the layers are closely situated. In the next section we present a way to overcome this challenge.

\subsection{Addressing the close evaluation problem (the BRIEF method)}
Achieving accurate computations near interfaces is a well-known challenge in boundary integral methods, addressed by a variety of techniques including Quadrature by Expansion \cite{klocknerQuadratureExpansionNew2013}, asymptotic expansions \cite{carvalhoAsymptotic2020}, and plane-wave density interpolation \cite{perezPlanewave2019}.
In earlier work \cite{cortesFastAccurateBoundary2024} on the multilayer scattering problem, we use the Boundary Regularized Integral Equation Formulation (BRIEF) method, which uses a modified solution representation to eliminate near-boundary errors \cite{sunBoundaryRegularizedIntegral2015}.
In the standard setting, the BRIEF method is applied to the representation formula (aka solutions in the layers).
However, in the background medium we can choose to evaluate the wave field behavior far enough off the boundary to neglect this type of error.
The complication specific to the multilayer transmission problem we mean to address arises when modeling thin cloaking layers.

More precisely, when boundaries $\Gamma_{j-1}$, $\Gamma_j$ are close, then one can find points $\mathbf{x}^b_{j-1} \in \Gamma_{j-1}$, $\mathbf{x}^b_j \in \Gamma_j$ such that $|\mathbf{x}^b_{j-1} - \mathbf{x}^b_j| \to 0$. Then the operators $\mathcal{S}_{j-1,j}, \mathcal{D}_{j-1,j},\mathcal{S}_{j,j},\mathcal{D}_{j-1,j}$ for $j =1, 
\dots N-2$ appearing in~\eqref{eq:boundaryrep} become nearly singular.
Because this error is introduced at the BIE system level, it reduces the accuracy of the wave field solution everywhere.

Let us present how to use BRIEF generally. Consider the solution of Helmholtz equation inside (or outside) a domain $\Omega$ with boundary $\Gamma$. Then at some point, the following combination of the single- and double- integral operators must be used: 
\begin{equation}\label{eq: sd}
\mathcal{D} [u](\mathbf{x}) - \mathcal{S} [\partial_n u](\mathbf{x})  .
\end{equation}
We assume \eqref{eq: sd} exhibits a nearly singular behavior when $|\mathbf{x} - \mathbf{x}^b| \to 0$. We introduce an auxiliary function $\psi$ and write
$$\mathcal{D} [u](\mathbf{x}) - \mathcal{S} [\partial_n u](\mathbf{x})  = \mathcal{D} [u - \psi](\mathbf{x}) - \mathcal{S} [\partial_n u - \partial_n \psi](\mathbf{x})  - \left(\mathcal{D} [\psi](\mathbf{x}) - \mathcal{S} [\partial_n \psi](\mathbf{x}) \right).$$
The goal is to define ad hoc $\psi$ such that:
\begin{itemize}
\item[(a)] $\mathcal{D} [u - \psi](\mathbf{x})$, $ \mathcal{S} [\partial_n u - \partial_n \psi](\mathbf{x})$ are regularized at \(\mathbf{x}^b\);

\item[(b)] \(\mathcal{D} [\psi](\mathbf{x}) - \mathcal{S} [\partial_n \psi](\mathbf{x})\) can be computed analytically (or with spectral accuracy).
\end{itemize}

One way to succeed is to construct $\psi$ solution of the Helmholtz equation in $\Omega$ such that $\psi(\mathbf{x}^b) = u (\mathbf{x}^b)$, $\partial_n \psi(\mathbf{x}^b) = \partial_n u (\mathbf{x}^b)$. With this choice (a) is achieved, as well as (b) via the Green's representation formula.  That is, it is well-known that solutions of the Helmholtz equation in a domain $\Omega$ satisfy
\begin{equation}
     \mathcal{D}[\psi](\mathbf{x}) - \mathcal{S}[\partial_n \psi](\mathbf{x}) = \begin{cases} \begin{aligned}
         -\psi(\mathbf{x}) & \quad \quad \mathbf{x} \in {\Omega} \\
 -\tfrac{1}{2}\psi(\mathbf{x})  & \quad \quad \mathbf{x} \in \Gamma, \\
 0 & \quad \quad \mathbf{x} \in \mathbb{R}^2 \setminus {\Omega} .
     \end{aligned} 
 \end{cases}
 \end{equation}
It is then just a matter of finding the appropriate domain $\Omega$ and auxiliary function $\psi$. We choose to construct $\psi$ as a linear combination of two solutions of Helmholtz equations (denoted $g,f$)
\[\psi(\mathbf{x)} = u(\mathbf{x}^b)g(\mathbf{x}) + \partial_n u(\mathbf{x}^b)f(\mathbf{x}),\]
where $g,f$ satisfy specific conditions at $\mathbf{x}^b$: $g(\mathbf{x}^b) = 1$, $n(\mathbf{x}^b) \cdot \nabla g(\mathbf{x}^b) = 0$, $f(\mathbf{x}^b) = 0$, $n(\mathbf{x}^b) \cdot \nabla f(\mathbf{x}^b) = 1$, with $n(\mathbf{x}^b)$ is the unit outward normal on $\Gamma$. By construction, (a) is guaranteed.

For the multilayered scattering problem, we utilize several auxiliary functions, one for each region $\Omega_j$ associated to wavenumber $k_j$, and the boundary integral equations must be modified wherever $|\mathbf{x}^b_{j-1} - \mathbf{x}^b_j| \to 0$. We introduce 
\begin{align*}
\psi_{j}(\mathbf{x)} &= u_{j,j}(\mathbf{x}^b_j)g_j(\mathbf{x}) + \partial_n u_{j,j}(\mathbf{x}^b_j)f_j(\mathbf{x}) &\quad \text{for } \mathbf{x} \in \Gamma_{j}, \quad j = 1, \dots N-2,
\end{align*}
with $$g_j(\mathbf{x}) = \cos\!\left(k_j\left[\mathbf{n}(\mathbf{x}^b_j) \cdot (\mathbf{x} - \mathbf{x}^b_j)\right]\right), \qquad f_j(\mathbf{x}) = \frac{1}{k_j}\sin\!\left(k_j\left[\mathbf{n}(\mathbf{x}^b_j) \cdot (\mathbf{x} - \mathbf{x}^b_j)\right]\right).$$
We now modify equations in \eqref{eq:boundaryrep} on $\Gamma_{j-1}$ as follows
\begin{equation*}\begin{aligned}
\frac{1}{2}u_j(\mathbf{x}^b_{j-1})  = & - D_{j-1, j}\left[u_{j-1, j-1}\right](\mathbf{x}^b_{j-1})  + \frac{\varepsilon_{j}}{\varepsilon_{j-1}} S_{j-1, j}\left[\partial_n u_{j-1,j-1} \right](\mathbf{x}^b_{j-1})   \\
& + \mathcal{D}_{j,j}\left[u_{j,j} -  \psi_{j}\right](\mathbf{x}^b_{j-1})  - \mathcal{S}_{j,j}\left[\partial_n u_{j,j} - \partial_n \psi_{j}\right](\mathbf{x}^b_{j-1}).  \\
\end{aligned}
\end{equation*}
Note that since ${\mathbf{x}}^b_{j-1} \in \Gamma_{j-1}$ is exterior to the domain enclosed by $\Gamma_j$, Green's representation formula applied to $\psi_j$ vanishes.
Then we modify equations in \eqref{eq:boundaryrep} on $\Gamma_{j}$: 
\begin{equation*}
\begin{aligned}
        \tfrac{1}{2}u_j(\mathbf{x}^b_j) 
            & + D_{j,j}[u_{j,j}](\mathbf{x}^b_j) - S_{j,j}[\partial_n u_{j,j}](\mathbf{x}^b_j) + \psi_j(\mathbf{x}^b_j)\\
        & -\mathcal{D}_{j-1,j}[u_{j-1,j-1} - \psi_j](\mathbf{x}^b_j) + \frac{\varepsilon_{j}}{\varepsilon_{j-1}}\mathcal{S}_{j-1,j}[\partial_n u_{j-1,j-1} - \partial_n\psi_j](\mathbf{x}^b_j). \\
\end{aligned}
\end{equation*}
Here, ${\mathbf{x}}^b_{j} \in \Gamma_{j}$ is interior to the domain enclosed by $\Gamma_{j-1}$, so Green's representation formula applied to $\psi_j$ yields the term $\psi_j(\mathbf{x}^b_j)$.

Although determining these modified expressions seems verbose, implementation of BRIEF within the boundary integral equations is in fact straightforward.
When present, the auxiliary function \(\psi_j(\mathbf{x})\) is explicitly known, and its associated trace and normal trace at \(\mathbf{x}^b\) are treated as unknowns within the BIE system.
Evaluation of the functions \(g_j, f_j\), and their normal derivatives is likewise direct.

Moreover, the assumptions adopted in our multilayer framework simplify the implementation further.
In particular, for concentric interfaces of similar geometries that differ only in \(r\), and which are discretized using the same \(M\) angular quadrature points \((\theta_m)_{m = 1, \dots, M}\), we assume that corresponding closest boundary points are aligned by their \(\theta\) values.
This eliminates the need to implement an additional method to determine these closest correspondences of points on different interfaces. 

\begin{figure}[!h]
    \centering    
    \includegraphics[width=0.8\textwidth]{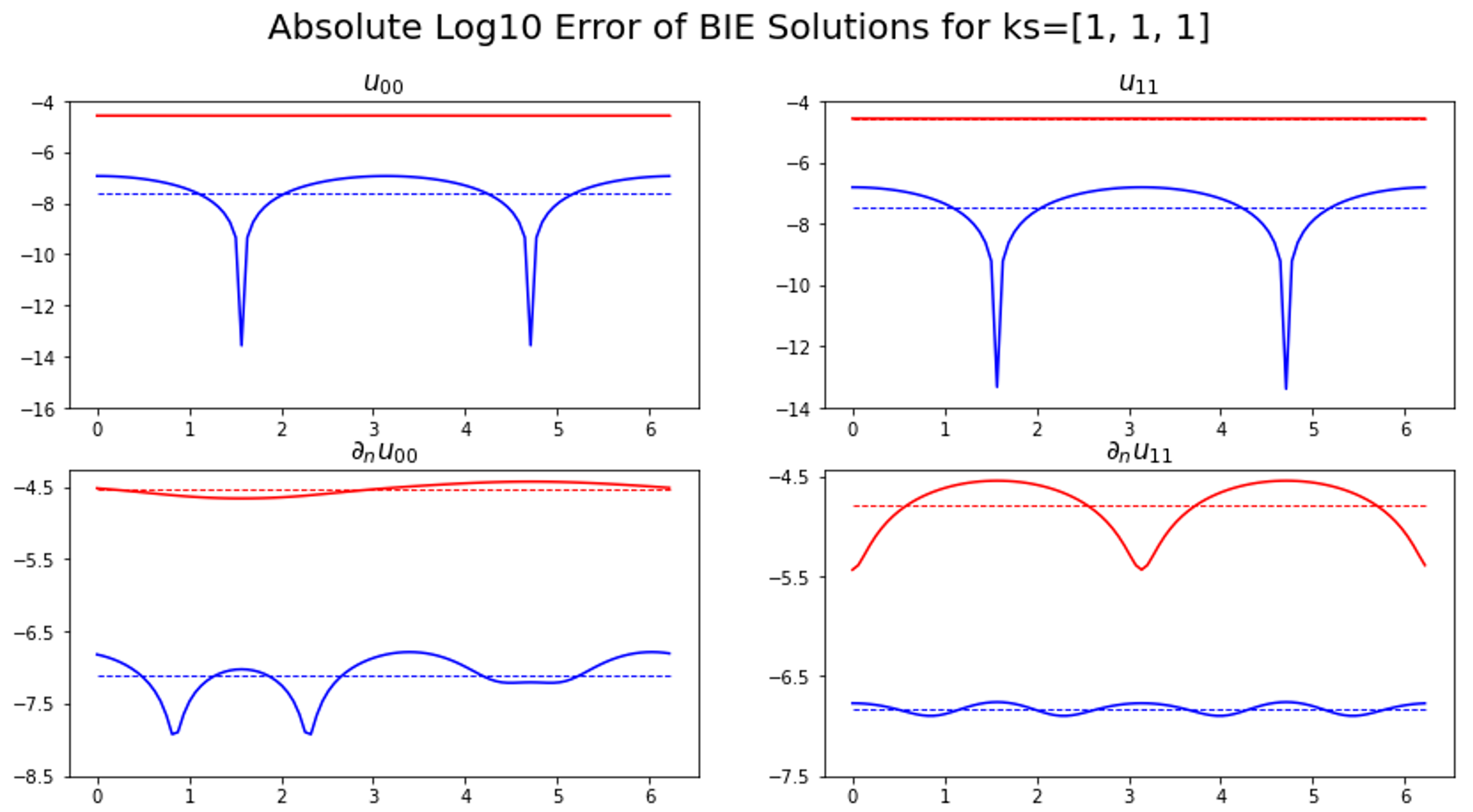} \includegraphics[width=0.8\textwidth]{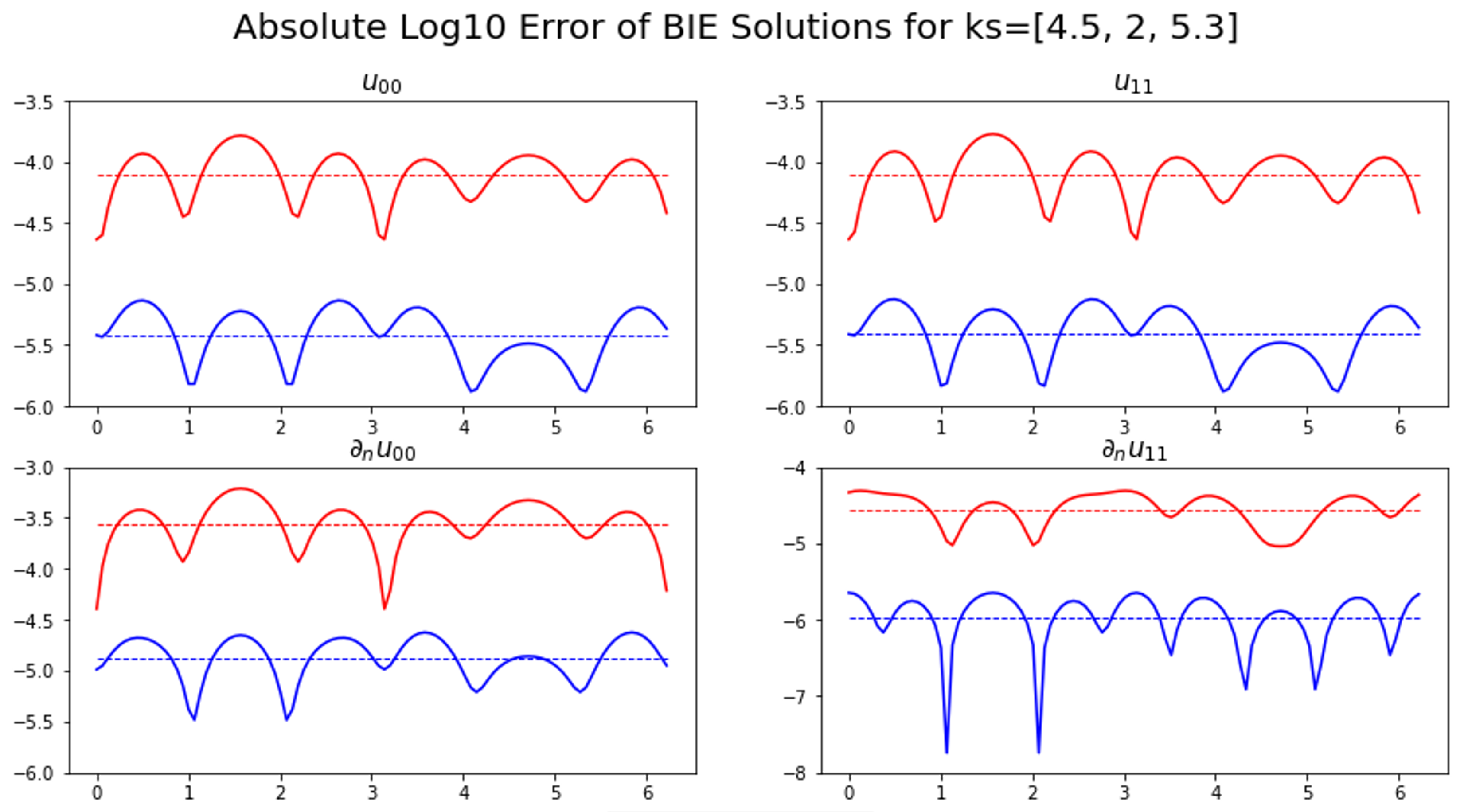}
    \caption{ Log plot of the absolute error of the boundary data \(u_{0,0}, u_{1,1}, \partial_n u_{0,0},\) and \(\partial_n u_{1,1}\) obtained with Kress Product Rule using \(M = 100\) equi-spaced quadrature points: without BRIEF (red), using BRIEF (blue).}
    \label{fig:BEBRIEFerrs}
\end{figure}

When the modified  system  is solved for these unknown traces and normal traces, we should observe an improvement in the accuracy of our boundary solutions.  This is seen in Figure~\ref{fig:BEBRIEFerrs}, where we show the boundary data errors before and after applying the BRIEF for a 3-layer forward problem with two circular boundaries. Both a homogeneous case with $(k_0,k_1,k_2) = (1,1,1)$ and a non-homogeneous case $(k_0,k_1,k_2) = (4.5,2,5.3)$ for radii \((r_0, r_1) = (1, 0.9)\) are represented. In both cases the method reduces the close-evaluation error by a significant margin. 

Although BRIEF does not eliminate the error entirely, it achieves accuracy comparable to what would otherwise require a substantially finer boundary discretization. This computational saving is modest for a single solve, but becomes significant when generating the large training and validation datasets required for training our data-driven neural network. BRIEF removes the uncertainty of the accuracy of our solutions for samples with cloaking layers characterized by very small \(\Delta r\), and this is reflected by the ability of our models to produce reliable results as shown in the following section.

\section{Results for trained neural network models}
\label{sec5}
\subsection{Choice of parameters and validation}
Having described the process for efficiently and accurately generating training and validation data, we now turn to the application of the neural network to the cloaking problem. Our goal is to evaluate how the architecture learns the mapping from inputs to outputs and to determine whether the models used to generate the outputs generalize well.  

Both the decoder and the combined encoder-decoder networks are trained using the Mean Squared Error (MSE) loss between the predicted and reference fields. For the decoder network, the objective is to learn the mapping from the input parameters to the output field. The combined encoder-decoder network is trained to approximate the identity mapping, where the input and output fields coincide. 

We note that each network is trained for a specific object geometry and material configuration $\varepsilon_{\rm{ob}}$, and for a chosen background medium $\varepsilon_{\rm{bg}}$. Additionally we fix the number of layers in the cloak. The generalizability of the framework lies in the architecture itself; the same encoder-decoder design and training procedure is applied across all configurations considered, with a corresponding dataset generated for each using the BEM pipeline described in Section \ref{sec4}. This makes the approach practical for new configurations, since accurate training data can be generated efficiently without the need for volumetric meshing.

In the following numerical experiments, we consider an object with characteristic radius \(r_{\rm{ob}} = 1\) and permittivity \(\varepsilon_{\rm{ob}} = 1\), embedded in a homogeneous background medium with permittivty \(\varepsilon_{\rm{bg}} = 4\).
We then analyze cloaking configurations, where we restrict our attention to a four-layer setting.
We generate samples by solving the forward problem for the object under varying layer parameters, recording the wave field over radii and permittivities satisfying predefined constraints.  For example, we choose that the layer boundaries and material parameters must satisfy

\begin{equation*}
\sum_{k=0}^3 \Delta r_k < 2.2 - r_{\rm{ob}} \quad  \text{and} \quad 0 < \varepsilon_j < 5\,\varepsilon_{\rm{bg}}, \quad j = 1, \ldots, 4.
\end{equation*} 

We also impose a minimum distance between consecutive radii, such that \(r_j - r_{j+1} > 0.01\).

While these material parameters and constraints are kept fixed, the geometry of the object and the configuration of the cloaking layers are varied in each case shown. Each dataset used consists of 10,000 samples, which we split into training, validation and test sets using 80-10-10 partition.  Larger datasets were also tested but there was not significant improvement in the loss behavior up to 30,000 samples.

\begin{figure}[!h]
    \centering
    \includegraphics[width=\linewidth]{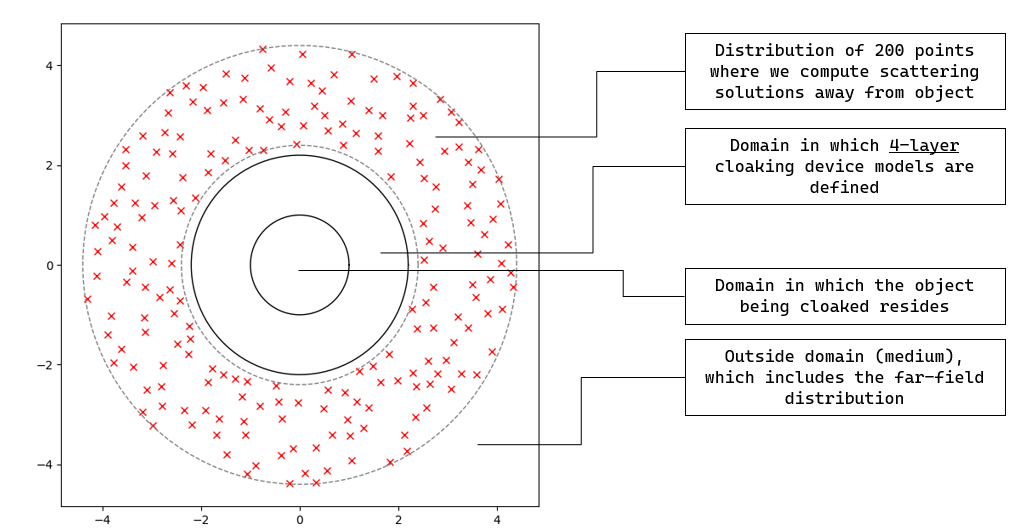}
    \caption{Sketch of the sample distribution and setting to generate dataset. The 200-point distribution on which wave field solution data is captured for the training of the multiple neural networks presented.  The sub-domains where the object and cloaking layers are defined are also shown.}
    \label{fig:distribution_pts}
\end{figure}

The sampling points of the wave field are constructed to evaluate the total field outside the cloaking region while avoiding numerical instability near the interfaces.  
Since we consider a four-layer cloak which extends up to a pre-defined maximum radius of \(r = 2.2\), all observation points are placed outside this region.
To balance accuracy and stability, we select a radial sampling domain spanning \(r \in [2.4, 4.4]\), ensuring a buffer exists between the outer close boundary and any evaluation points to avoid close evaluation errors from the BEM in the representation formula.  
Within this annular region, points are distributed uniformly in \(\theta \in [0, 2\pi)\) with small random angular shifts to avoid systematic alignment with geometric features of the configuration and reduce directional clustering.
The resulting distribution (shown in Figure \ref{fig:distribution_pts}) consists of 200 points randomly sampled in \(r\), which corresponds to an effective density of 5 points per unit area.
We note that this yields approximately 7 points per wavelength resolution with respect to the wavelength of the background medium, \(\lambda_{\rm{bg}}= \pi\), which is a moderate-use resolution for wave propagation in this setting \cite{CCCT2026}.

Including points closer to the cloak may improve the network's ability to resolve the influence of the layered structure on the nearby scattered field, though it also increases the difficulty of the learning task.  Smaller and larger point sets were also evaluated, but the 200-point distribution was the smallest that achieved stable and accurate performance.  Larger point distributions did not yield sufficient improvement to justify the additional computational cost.  The overall difficulty of the problem also depends on the permittivity ratio \(\varepsilon_{\mathrm{ratio}} := \varepsilon_{\mathrm{bg}}/\varepsilon_{\mathrm{ob}}\), which controls the intensity of the scattered field produced by the object. 

The same neural network architecture is employed in all settings so we may see how well it generalizes across the range of object and cloak geometries considered.  Nevertheless, more substantial changes in problem setup, such as varying the object radius or number of cloaking layers, may require corresponding modifications to dataset size, sampling distribution, and potentially other training hyperparameters.

 Model checkpoints are saved whenever the average validation MSE loss improves relative to the best value observed in previous training passes (epochs).  The final model corresponds to the network achieving the lowest validation MSE loss, ensuring that the model selection prioritizes generalization rather than simply fitting the training data.

\begin{figure}[!h]
    \centering
    \begin{subfigure}{0.5\linewidth}
        \centering
        \includegraphics[width=\linewidth]{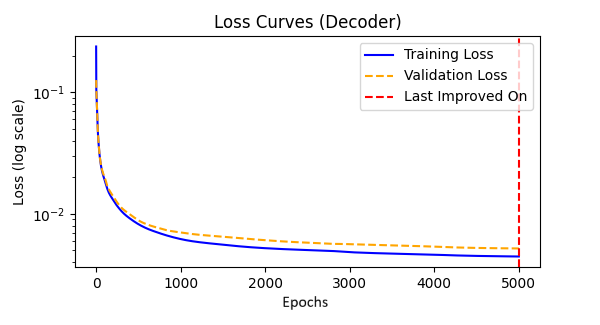}
        \caption{Decoder network}
        \label{fig:decoder_losses}
    \end{subfigure}%
    \begin{subfigure}{0.5\linewidth}
        \centering
        \includegraphics[width=\linewidth]{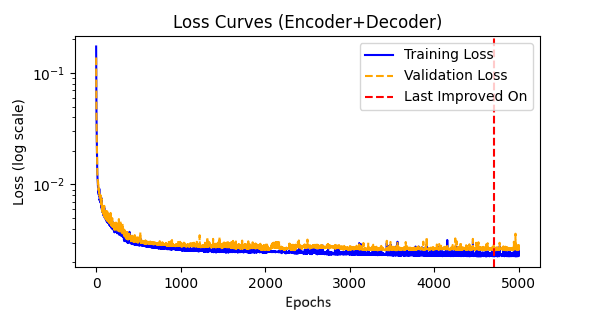}
        \caption{Combined encoder-decoder network}
        \label{fig:combined_losses}
    \end{subfigure}
    
    \caption{Training and validation MSE loss curves for the simple circular-object case with $r_{\rm{ob}}=1$, $\varepsilon_{\rm{ob}}=1$, and $\varepsilon_{\rm{bg}}=4$ for the decoder network and the combined encoder-decoder network.}
    \label{fig:loss_comparison} 
\end{figure}

In Figure \ref{fig:decoder_losses}, we show the training and validation MSE loss curves for the decoder architecture described in the previous subsection.

The overall smoothness of the loss curves indicates a well-behaved optimization process that largely avoids instabilities or divergence, which could otherwise lead to inconsistent predictions.  
This decoder architecture is sufficiently stable to be used within the full encoder-decoder framework to predict the wave field with satisfactory accuracy. 
This example is representative of the general training behavior observed across all problem settings and model variants studied in the remainder of this paper. 
We improve the validation loss up to the 5000 epochs used for training, which indicates the model may even be further improved given additional time.

The MSE loss curves for the same case using the combined encoder-decoder architecture are shown in Figure \ref{fig:combined_losses}.  
Utilizing the pre-trained decoder model, the network trains effectively with no apparent underfitting or overfitting.  
In this particular case, the combined network exhibits slightly faster convergence than the decoder model alone, although this behavior is not consistent and it sometimes converges slower. 
Overall, fluctuations in the loss are minimal; while minor variability is present, the overall loss trend remains consistently downward.

\begin{figure}[!b]
    \centering
    \includegraphics[width=\linewidth]{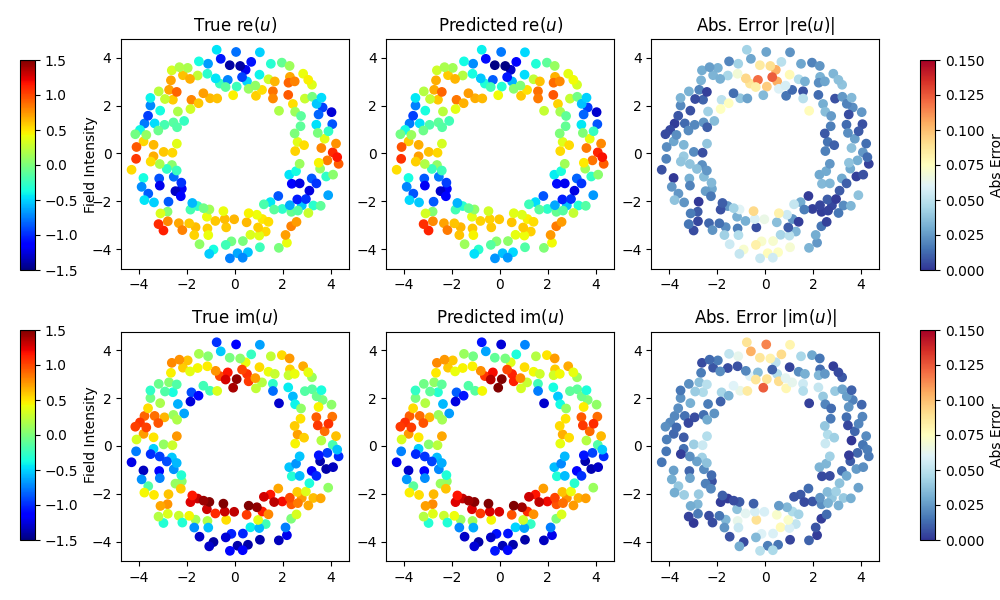}
    \caption{Evaluation of the decoder's output for one test sample.  The top row corresponds to the real component of the wave field solutions, and the bottom row to the imaginary component.  The 200-point distribution used to generate wave field data is where accuracy is evaluated.}
    \label{fig:decoder_sample}
\end{figure}

We further evaluate the accuracy of the fully trained decoder model by computing the absolute error between test samples and their corresponding predictions.
We look at one sample case in Figure \ref{fig:decoder_sample}, which illustrates the accuracy of the model across most cases.
Visually, the true and predicted real and imaginary components of the wave field solutions for the data distribution correspond well.
The pointwise absolute errors are minimal, with only a small number of localized regions exhibiting slightly elevated yet still acceptable error.
These results show that the decoder model has been trained to a high degree of accuracy, which is essential for enabling effective training of the encoder as well.

\begin{figure}[!t]
    \centering
    \includegraphics[width=\linewidth]{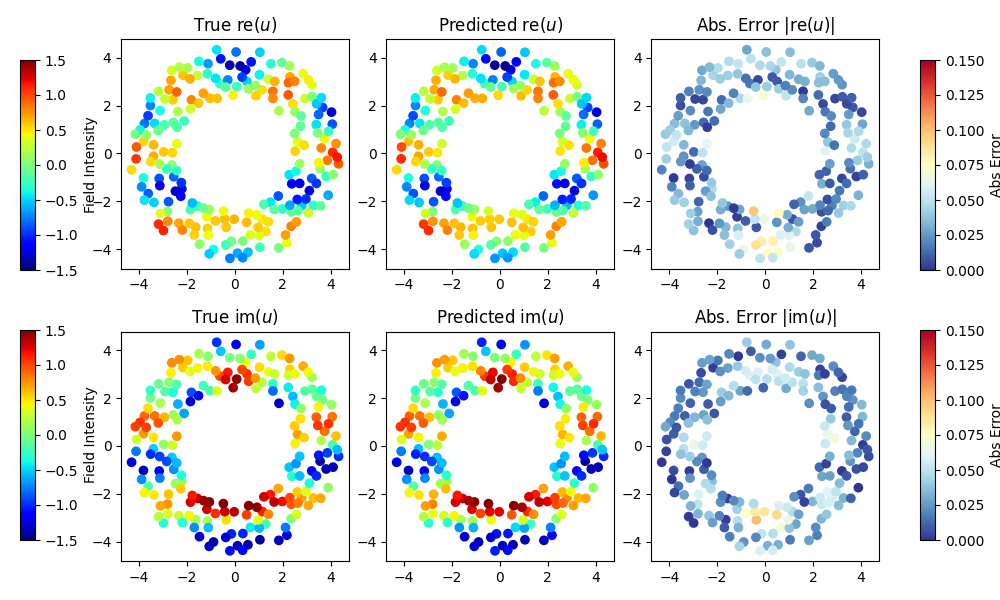}
    \caption{Evaluation of the encoder-decoder's output for the test sample in Figure \ref{fig:decoder_sample}.}
    \label{fig:encoder_sample}
\end{figure}

We also evaluate the absolute errors for the combined encoder-decoder by comparing input to output wave field solutions directly.  We examine the same sample from the previous figure in Figure \ref{fig:encoder_sample}. The highest magnitude of error is lower than what was observed in the decoder, which tells us that the combined network is well trained to produce approximately the same wave field from the parameters output by the encoder in the model's latent layer.  This also shows that the general performance of the model for predicting the wave field is improved with this combined network construction. 

\subsection{Evaluation of a Cloak with a Circular Object}

Having examined the loss curves and outputs of the models, we now apply the trained encoder to generate cloaking design parameters. The incident wave field for \(\varepsilon_{\rm{bg}} = 4\) is fed into the encoder, which outputs the cloaking layer parameters needed to mitigate scattering.

To quantify the performance of the cloak, we compute the radar cross section (RCS), which measures the visibility of the object to an incoming wave: a larger RCS indicates stronger scattering and greater detectability. The RCS is evaluated on a circle of radius 
\begin{equation*}
    R = m \lambda_{\rm{bg}} = m\frac{2\pi}{k_{\rm{bg}}},
\end{equation*}
where \(\lambda_{\rm{bg}}\) is the wavelength in the background medium and $m$ is the number of wavelengths at which we measure. Generally \(1 \leq m \leq 3\) is used to evaluate the near-field behavior of the solution and \(m \geq 5\) to approximate it in the far-field. We will use \(m = 5\). The RCS in 2D is then defined as 
\begin{equation}
    \sigma(\theta) = 2\pi R \frac{|u_{\rm{bg}}(R, \theta) - \uincbg(R, \theta)|^2}{|\uinc(R, \theta)|^2}.
    \label{eq:rcs}
\end{equation}

We compute both the object RCS with no cloak present, and the cloak RCS using the parameters output by the encoder, and compare them visually.

\begin{figure}[!h]
    \centering
    \includegraphics[width=0.85\linewidth]{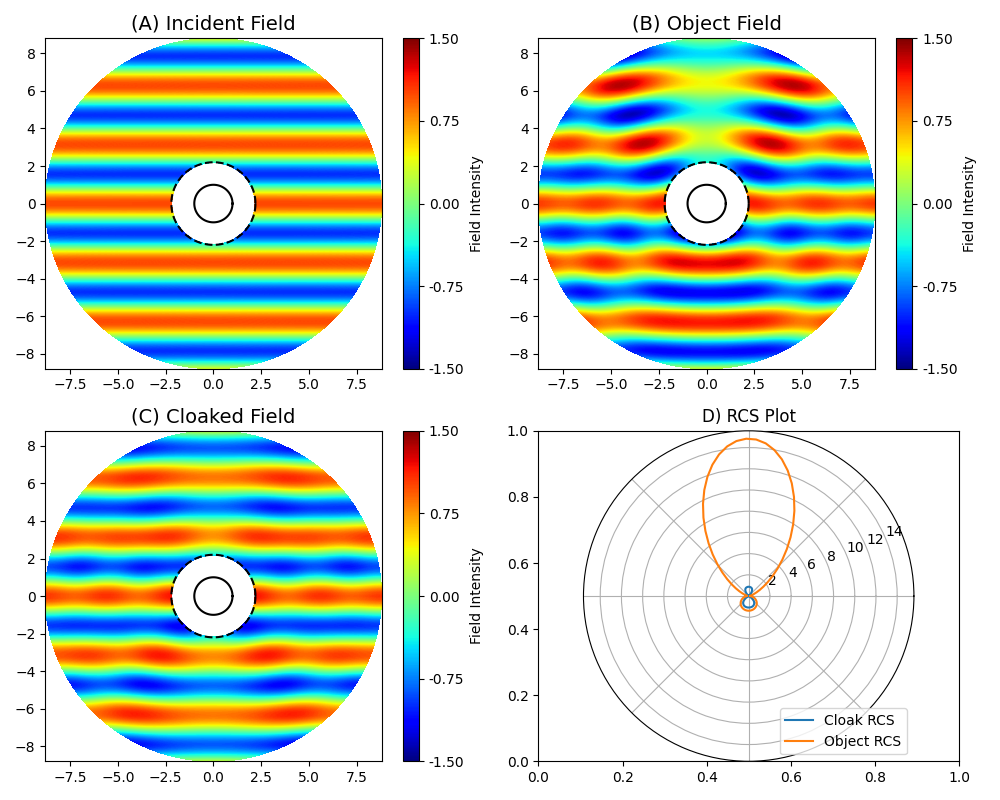}
    \caption{Visualization of the cloaking performance of the neural network.  Only the real part of each field is shown on the contour plots.  (A) Incident field \(\uincbg\), showing the field that impinges on the object. (B) Object field $u_{\rm{bg}}$ without any cloak. (C) Cloaked field $u_{\rm{bg}}$ obtained using the neural-network-generated parameters which minimize scattering. (D) Radar Cross Section (RCS) plot comparing the object (orange) and the cloaked object (blue) which is almost entirely minimized.}
    \label{fig:circ_circ_sample}
\end{figure}

The incident field is shown in Figure \ref{fig:circ_circ_sample}(A), the uncloaked object field is displayed in Figure \ref{fig:circ_circ_sample}(B) and the cloaked in Figure \ref{fig:circ_circ_sample}(C), which closely reproduces the incident field.  The corresponding RCS curves are plotted in Figure \ref{fig:circ_circ_sample}(D). 

\begin{figure}[!h]
\centering
\begin{minipage}{0.45\linewidth}
\centering
\includegraphics[width=0.75\linewidth]{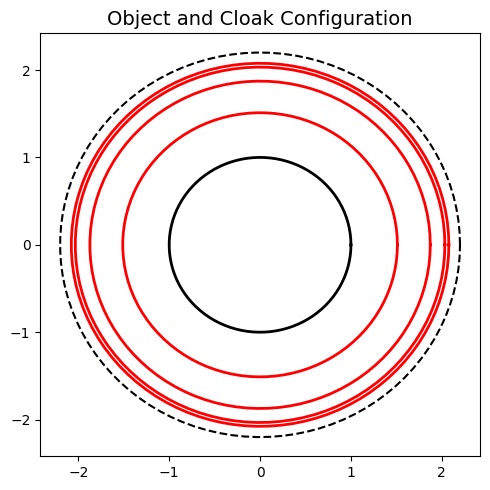}
\label{fig:object_cloak_config}
\end{minipage}%
\begin{minipage}{0.45\linewidth}
\begin{tabular}{|c|c|}
\hline
$\Delta\mathbf{r}$ & $\mathbf{e}$ \\
\hline \hline
0.0436 & 1.072 \\ \hline
0.2058 & 2.114 \\ \hline
0.3617 & 1.303 \\ \hline
0.5112 & 1.469 \\
\hline
\end{tabular}
\end{minipage}
\caption{Cloaking parameters produced by the neural network for the simple circular-object case with $r_{\rm{ob}}=1$, $\varepsilon_{\rm{ob}}=1$, and $\varepsilon_{\rm{bg}}=4$. The left panel shows the object interface in black and cloak interfaces in red, while the right panel lists the corresponding layer radii and permittivities.}
\label{fig:circ_cloak_params}
\end{figure}

These results demonstrate that the proposed architecture is capable of producing a highly effective cloaking design for the circular-object case.  The cloaking device and corresponding layer radii and permittivities are shown in Figure~\ref{fig:circ_cloak_params}.
The loss behaviors observed in the models trained for this case is representative of all models trained in this work, and comparably effective cloaking results are obtained across all configurations considered.
In the additional examples explored in the next sections, the RCS in \eqref{eq:rcs} will serve as the primary quantitative metric for assessing cloak performance, enabling systematic comparison across different object and layer geometries.

We make a couple of remarks:
\begin{itemize}
    \item The output cloak by the neural network in Figure \ref{fig:circ_cloak_params} considers rather thin layers. It is then essential to ensure accurate computation of the boundary data in such configuration.
    \item In the case of circular interfaces, one could directly consider the analytic solution to generate data and train the network. However, this approach remains limited. In the next section we test our methodology on other geometries, where no analytic solution is available.
\end{itemize}

\subsection{Cloaking a Star-shaped Object}

Having demonstrated effective cloaking for circular objects, we now consider geometries with non-circular boundaries.  As a first step, we introduce small perturbations of the circular object and layer boundaries, which we refer to as  \textit{star-shaped}, parameterized by
\begin{equation}\label{eq:star}
\begin{aligned}
x(\theta) &= \big(r_j + A\cos(n \theta)\big)\cos \theta, \\
y(\theta) &= \big(r_j + A \cos(n \theta)\big)\sin \theta,
\end{aligned}
\end{equation}
where \(n\) is the number of star points and \(A \geq 0\) controls their relative size with respect to \(r_j\). In the following examples, we choose \(n = 10\) and \(A = 0.05\), only varying \(r_j\).  

\begin{figure}[!h]
    \centering
    \begin{minipage}{0.65\linewidth}
    \includegraphics[width=0.5\linewidth]{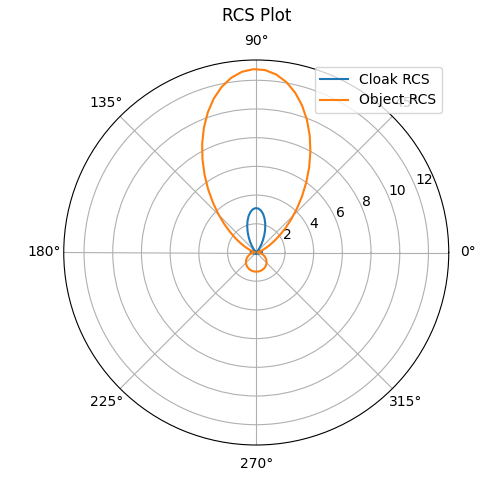} \includegraphics[width=0.45\linewidth]{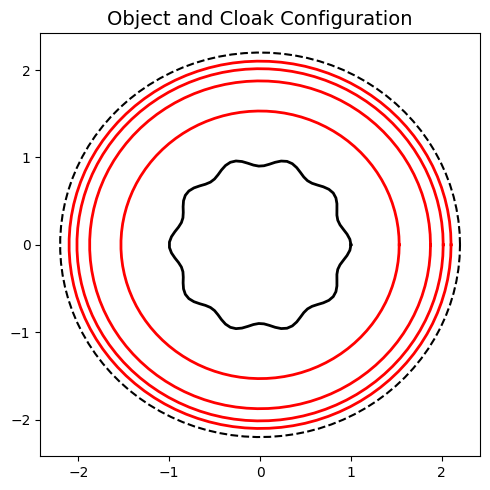} 
    \label{fig:circ_star_config}
    \end{minipage}
    \hspace{0.01\linewidth}
    \begin{minipage}{0.3\linewidth}
    \begin{tabular}{|c|c|}
    \hline
    $\Delta\mathbf{r}$ & $\mathbf{e}$ \\
    \hline \hline
    0.0865 & 12.48 \\ \hline
    0.1398 & 1.075 \\ \hline
    0.3443 & 10.35\\ \hline
    0.5310 & 4.197 \\
    \hline
    \end{tabular}
    \end{minipage} 
    \caption{Cloaking parameters produced by the neural network for the star-object case with \textit{circular layers} for $r_{\rm{ob}}=1$, $\varepsilon_{\rm{ob}}=1$, and $\varepsilon_{\rm{bg}}=4$. The RCS plot, the object and cloaking layers, and the corresponding layer radii and permittivities are shown.}
    \label{fig:circ-star-fig}
    \end{figure}

For non-circular objects, an immediate design question is how the cloaking layers should be shaped. Cylindrical or spherical layers are often preferred for their simplicity of fabrication, accessibility to analytic solutions, and ease of coordinate transformation \cite{pendry2006, InvisibilityCloakingCoordinate2009}.
However, one may ask whether \textit{object-fitted} layers, whose boundaries follow the star-shaped object boundary, would yield superior scattering reduction.
Our neural network can generate parameters for either choice of layer geometry, enabling a direct comparison via RCS evaluations.

Figure \ref{fig:circ-star-fig} shows the results when \textit{circular layers} are used to cloak the star-shaped object, with $r_{\rm{ob}}=1$, $\varepsilon_{\rm{ob}}=1$, and $\varepsilon_{\rm{bg}}=4$, the same parameters as the circular case, for which our sampling distribution is well validated.  The encoder-decoder produces an effective cloak, though with somewhat reduced performance compared to the circular object case.

\begin{figure}[!t]
\centering
\begin{minipage}{0.65\linewidth}
\includegraphics[width=0.5\linewidth]{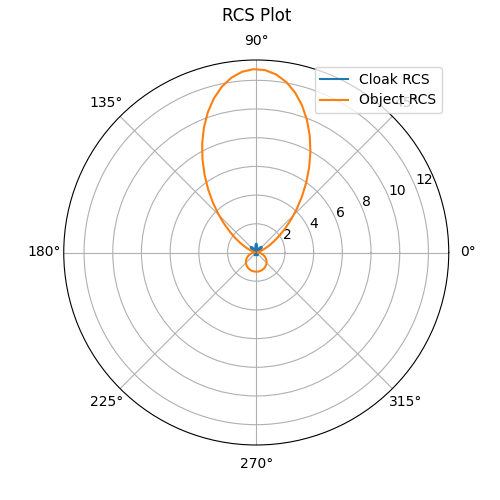} \includegraphics[width=0.45\linewidth]{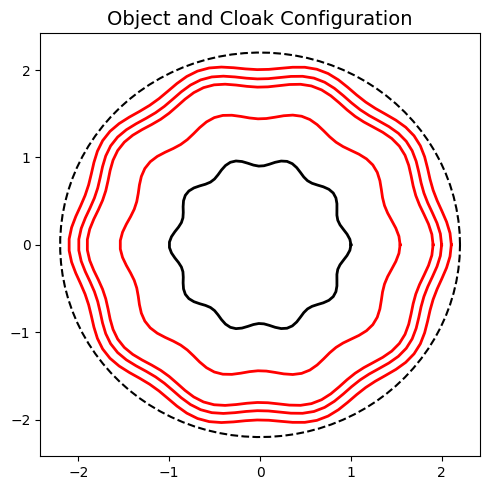} 
\label{fig:star_star_config}
\end{minipage}%
\begin{minipage}{0.3\linewidth}
\begin{tabular}{|c|c|}
\hline
 $\Delta \mathbf{r}$ & $\mathbf{e}$ \\
\hline \hline
0.1065  & 15.63 \\ \hline
0.0939 & 1.171 \\ \hline
0.3628 & 7.409 \\ \hline
0.4909 & 5.154 \\
\hline
\end{tabular}
\end{minipage}
\caption{Cloaking parameters produced by the neural network for the star-object case with \textit{star-shaped layers} corresponding to the problem in Figure \ref{fig:circ-star-fig}. }
\label{fig:star-star-fig}
\end{figure}

Figure \ref{fig:star-star-fig} presents the results when both the object and layer boundaries are star-shaped. Despite very similar predicted parameters, the RCS shows a clear improvement over the circular-layer case, with the object-fitted cloak achieving scattering reduction comparable to, or better than, what was obtained for the simpler circular geometry. This confirms that the same general architecture can handle different boundary shape configurations, and that object-fitted layers can indeed yield more effective cloaking.

Whether the added fabrication complexity of star-shaped layers is justified by the improvement in cloaking performance is ultimately a practical design decision. The neural network provides a fast means to explore such trade-offs, and the use of BEM with BRIEF enables accurate evaluation of these configurations without the computational overhead of volumetric methods such as finite element or finite difference methods.

These results illustrate the key ideas well, but the star-shaped geometry remains a relatively mild perturbation of the circular case. In particular, all objects considered so far are radially symmetric, with wave field responses that do not strongly depend on their orientation relative to the incident field. In the next subsection, we consider kite-shaped objects, where geometric irregularity and orientation both significantly affect the scattering behavior, providing a more demanding test of the encoder-decoder's generalization.

\subsection{Cloaking Kite-Shaped Objects}

The kite-shaped boundary is parameterized as a perturbation of a circle of base radius \(r_j\):
\begin{equation}\label{eq: kite}
\begin{aligned}
    x(\theta) &= r_j\big( \cos \theta + A \cos(n\theta)\big) - Ar_j,  \\
    y(\theta) &= r_j \sin \theta,
\end{aligned}
\end{equation}

where \(A \geq 0\) controls the amplitude of the perturbation.  In the following examples, we choose \(A = 0.5\).  
The kite is symmetric with respect to the \(y\)-axis, and we examine two orientations relative to the incident field direction: one in which the convex side faces the incoming wave, and one in which the concave side does.  To produce a significant object RCS, we take $r_{\rm{ob}}=0.5$, $\varepsilon_{\rm{ob}}=36$, and $\varepsilon_{\rm{bg}}=4$. 

As before, we compare results for circular and object-fitted (kite-shaped) cloaking layers. While the approach has proven effective for the geometries considered so far, cloaking a kite-shaped object is a considerably more challenging problem due to its geometric irregularity and orientation dependence. It is precisely for such objects that a flexible, data-driven approach is most valuable.

\begin{figure}[!t]
\centering
\begin{minipage}{0.65\linewidth}
\includegraphics[width=0.5\linewidth]{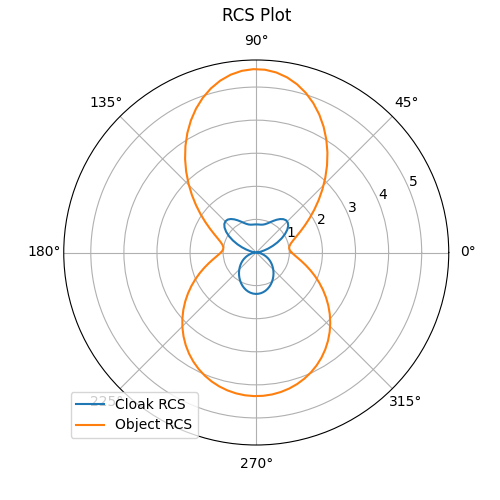} \includegraphics[width=0.45\linewidth]{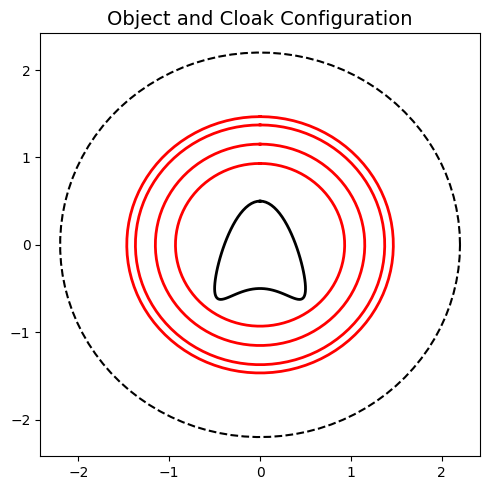} 
\label{fig:circ_kite_pos_config}
\end{minipage}
\hspace{0.01\linewidth}
\begin{minipage}{0.3\linewidth}
\begin{tabular}{|c|c|}
\hline
$\Delta\mathbf{r}$ & $\mathbf{e}$ \\
\hline \hline
0.0949 & 9.727 \\ \hline
0.2194 &  8.550\\ \hline
0.2213 & 3.027 \\ \hline
0.4309 & 1.029 \\
\hline
\end{tabular}
\end{minipage} 
\caption{Cloaking parameters produced by the neural network for the \textit{convex-facing} kite-object case with \textit{circular layers} for $r_{\rm{ob}}=0.5$, $\varepsilon_{\rm{ob}}=36$, and $\varepsilon_{\rm{bg}}=4$.}
\label{fig:circ-kite-fig}
\end{figure}

We begin with the less challenging orientation of the kite to cloak, where the convex side of the kite faces the incident field.  The results from our neural network for \textit{circular layers} cloaking the kite-shaped object are given in Figure \ref{fig:circ-kite-fig}.  An interesting observation emerges from the predicted parameters: in typical metamaterial cloaking designs, the layer permittivities alternate between higher and lower values across consecutive interfaces. This alternating pattern is seen in all previous examples. For this configuration, however, the permittivities decrease monotonically across the layers. This is an atypical construction, yet one that the RCS confirms to be effective.

\begin{figure}[!t]
\centering
\begin{minipage}{0.65\linewidth}
\includegraphics[width=0.5\linewidth]{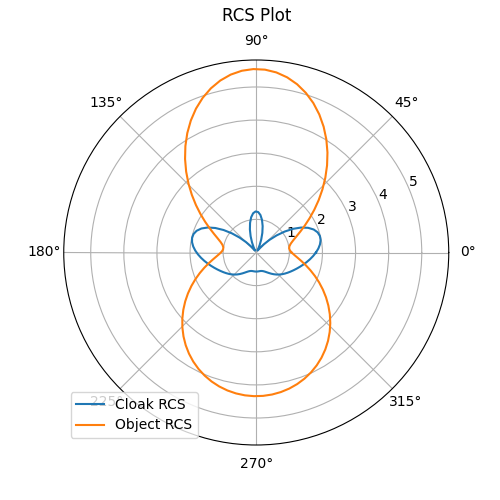} \includegraphics[width=0.45\linewidth]{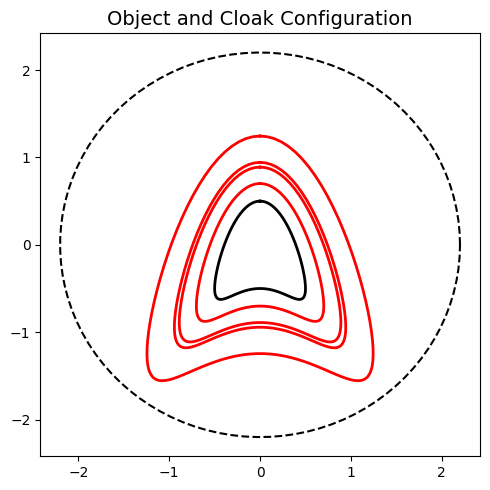} 
\label{fig:kite_kite_pos_config}
\end{minipage}
\hspace{0.01\linewidth} 
\begin{minipage}{0.3\linewidth}
\begin{tabular}{|c|c|c|}
\hline
$\Delta\mathbf{r}$ & $\mathbf{e}$ \\
\hline \hline
0.3123 & 1.862 \\ \hline
0.0721 & 12.02 \\ \hline
0.1832 & 17.66 \\ \hline
0.1810 & 35.97 \\
\hline
\end{tabular}
\end{minipage}
\caption{Cloaking parameters produced by the neural network for the kite-object case with \textit{kite-shaped layers} corresponding to the problem in Figure \ref{fig:circ-kite-fig}.}
\label{fig:kite-kite-fig}
\end{figure}

When \textit{kite-shaped layers} are used instead (Figure~\ref{fig:kite-kite-fig}), the cloaking is not fully effective overall: while scattering is reduced in the direction of the incident field, it is slightly increased in the cross directions. This suggests that object-fitted kite-shaped layers are particularly effective at suppressing scattering in the direction they are oriented, which may be advantageous in applications where directional scattering reduction is the primary objective. 

\begin{figure}[!t]
\centering
\begin{minipage}{0.65\linewidth}
\includegraphics[width=0.5\linewidth]{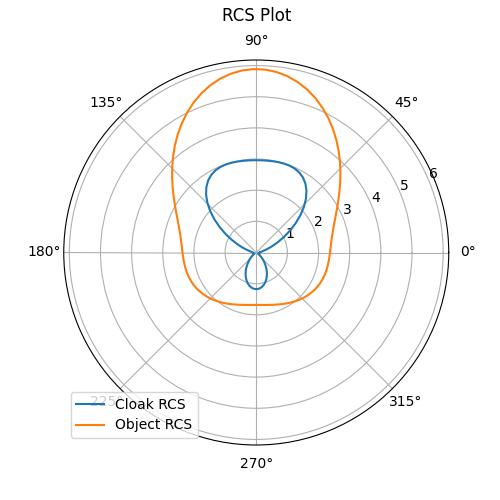} \includegraphics[width=0.45\linewidth]{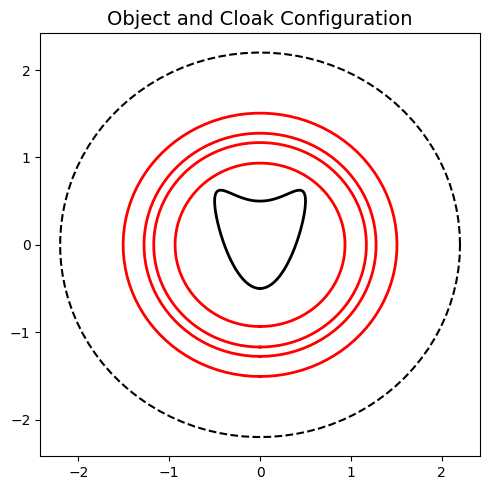} 
\label{fig:circ_kite_neg_config}
\end{minipage}
\hspace{0.01\linewidth}
\begin{minipage}{0.3\linewidth}
\begin{tabular}{|c|c|}
\hline
$\Delta\mathbf{r}$ & $\mathbf{e}$ \\
\hline \hline
0.2097 & 9.379 \\ \hline
0.0943 & 5.254 \\ \hline
0.2631 & 1.562 \\ \hline
0.4078 & 0.980 \\
\hline
\end{tabular}
\end{minipage}

\caption{Cloaking parameters produced by the neural network for the \textit{concave-facing} kite-object case with \textit{circular layers} for $r_{\rm{ob}}=0.5$, $\varepsilon_{\rm{ob}}=36$, and $\varepsilon_{\rm{bg}}=4$.}
\label{fig:circ-kite-fig-2}
\end{figure}

\begin{figure}[!t]
\centering
\begin{minipage}{0.65\linewidth}
\includegraphics[width=0.5\linewidth]{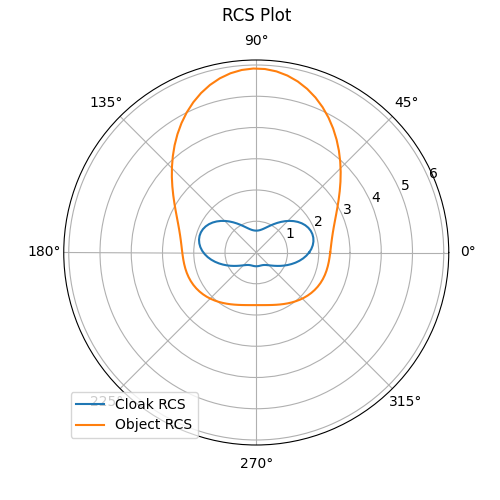} \includegraphics[width=0.45\linewidth]{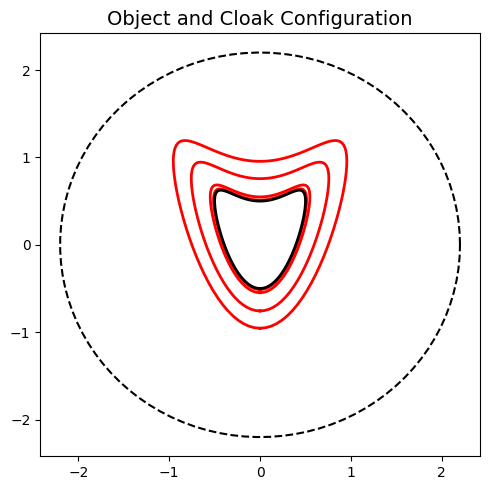} 
\label{fig:kite_kite_neg_config}
\end{minipage}
\hspace{0.01\linewidth} 
\begin{minipage}{0.3\linewidth}
\begin{tabular}{|c|c|}
\hline
$\Delta\mathbf{r}$ & $\mathbf{e}$ \\
\hline \hline
0.1980 & 1.878 \\ \hline
0.2086 & 0.923 \\ \hline
0.0362 & 2.157 \\ \hline
0.0119 & 1.334 \\
\hline
\end{tabular}
\end{minipage}
\caption{Cloaking parameters produced by the neural network for the kite-object case with \textit{kite-shaped layers} corresponding to the problem in Figure \ref{fig:circ-kite-fig-2}.}
\label{fig:kite-kite-fig-2}
\end{figure}

Cloaking the kite-shaped object when its concave side faces the incident field is expected to be more challenging.  In this orientation, we may conceive that a cloak composed of \textit{circular layers} could outperform the object-fitted cloak with \textit{kite-shaped layers}.  The neural networks allows us to directly gain insight into this question. We use the same object and background parameters as in the previous kite-shaped example, modifying only the object orientation.  Figure \ref{fig:circ-kite-fig-2} shows the neural-network-predicted cloaking configuration and RCS for \textit{circular layers}, while Figure \ref{fig:kite-kite-fig-2} presents the corresponding results for \textit{kite-shaped layers}.  

In both cases, the neural network identifies a viable cloaking design.  Despite the unfavorable concave-facing orientation, the object-fitted cloak again yields superior reduction of the object's scattered field.
Most of the scattering in the incident field direction is reduced, although residual scattering remains in the \(\pi\) and \(2\pi\) side directions.
The results further support the observation that object-fitted cloaking configurations tend to outperform simpler layer geometries.
More broadly, they demonstrate that the neural network architecture can be readily extended to assess more complex object shapes for which analytic solutions are not available.  

In this context, the full pipeline including the BEM used to generate training data is particularly valuable for modeling cloaking designs.
Many of the cloak configurations considered in this work involve very thin layers, and without the use of the BRIEF method, samples intended to capture the effect of such thin layers on wave propagation would be unreliable.
These findings collectively underline the robustness of the proposed framework and its considered applicability to increasingly complex cloaking problems.

\section{Conclusions}
\label{sec6}
We have presented a data-driven framework for the design and evaluation of optical cloaking devices, based on an encoder-decoder neural network trained on accurate BEM-generated scattering data. The framework enables systematic inverse determination of cloaking layer parameters and direct comparison of different cloaking strategies across a range of object geometries.

A central technical contribution is the integration of the BRIEF method into the data generation pipeline, which enables accurate BEM solutions for thin-layer configurations without the need for prohibitively fine boundary discretizations. This extends the applicability of the framework well beyond simple circular geometries. The neural network consistently identified effective cloaking designs for circular, star-shaped, and kite-shaped objects, including configurations where orientation and geometric concavity significantly influenced the scattering behavior.

A consistent finding across all cases is that object-fitted cloaking layers outperform circular layers in terms of scattering reduction, as measured by the RCS. This challenges the common assumption in optical cloaking design that geometrically simpler, symmetric shells are universally optimal. The RCS provides a physically meaningful and consistent metric for quantifying these differences, making it straightforward to assess trade-offs between geometric complexity and cloaking effectiveness.

Several directions remain open for future work.  First, we restricted attention to 2D wave propagation, but this framework can be extended to 3D where the BEM would make data generation  highly efficient.
Additionally, the present framework is demonstrated for time-harmonic propagation of electromagnetic waves, but the same methodology can be extended to other wave phenomena including acoustic and elastic scattering with appropriate modifications to the governing equations and boundary conditions.
Another natural extension concerns the complexity of the cloaking design itself.  In the present work, a fixed number of layers is assumed to enable a controlled comparison across geometries.  Future work could consider variable-depth or adaptive layer constructions, where the number of layers is treated as part of the design problem, with the goal of achieving comparable cloaking performance using simpler configurations.
This would allow the framework to explore trade-offs between geometric complexity and scattering reduction.
Additionally, the current approach uses separate models for circular and object-fitted configurations, but a unified training strategy could enable the network to directly infer the optimal layer geometry for a given object. Such developments would further improve the generalization of this framework, establishing it as a stronger general-purpose tool for inverse scattering designs.

\section{Acknowledgments}
\begin{itemize}
    \item C.~Tsogka acknowledges support by the Air Force Office of
Scientific Research (AFOSR FA9550-23-1-0352 and FA9550-24-1-0191). 
    \item E.~Cortes acknowledges support by NSF Grant DMS-1840265.
\end{itemize}

\bibliographystyle{elsarticle-num}
\bibliography{main}

@article{Ahmed2021,
  title = {Deterministic and probabilistic deep learning models for inverse design of broadband acoustic cloak},
  author = {Ahmed, Waqas W. and Farhat, Mohamed and Zhang, Xiangliang and Wu, Ying},
  journal = {Phys. Rev. Res.},
  volume = {3},
  issue = {1},
  pages = {013142},
  numpages = {8},
  year = {2021},
  month = {Feb},
  publisher = {American Physical Society},
  doi = {10.1103/PhysRevResearch.3.013142},
}

@article{alu2005,
  title = {Achieving Transparency with Plasmonic and Metamaterial Coatings},
  author = {Al{\`u}, Andrea and Engheta, Nader},
  year = 2005,
  month = jul,
  journal = {Physical Review E},
  volume = {72},
  number = {1},
  pages = {016623},
  publisher = {American Physical Society},
  doi = {10.1103/PhysRevE.72.016623},
}

@article{cai2007,
  title = {Optical Cloaking with Metamaterials},
  author = {Cai, Wenshan and Chettiar, Uday K. and Kildishev, Alexander V. and Shalaev, Vladimir M.},
  date = {2007-04},
  journaltitle = {Nature Photonics},
  shortjournal = {Nature Photon},
  volume = {1},
  number = {4},
  pages = {224--227},
  publisher = {Nature Publishing Group},
  issn = {1749-4893},
  doi = {10.1038/nphoton.2007.28},
}

@online{dubey2022,
  title = {Activation {{Functions}} in {{Deep Learning}}: {{A Comprehensive Survey}} and {{Benchmark}}},
  shorttitle = {Activation {{Functions}} in {{Deep Learning}}},
  author = {Dubey, Shiv Ram and Singh, Satish Kumar and Chaudhuri, Bidyut Baran},
  date = {2022-06-28},
  eprint = {2109.14545},
  eprinttype = {arXiv},
  eprintclass = {cs},
  doi = {10.48550/arXiv.2109.14545},
}

@article{Kiarashinejad2020,
  title = {Deep Learning Approach Based on Dimensionality Reduction for Designing Electromagnetic Nanostructures},
  author = {Kiarashinejad, Yashar and Abdollahramezani, Sajjad and Adibi, Ali},
  date = {2020-02-04},
  journaltitle = {npj Computational Materials},
  shortjournal = {npj Comput Mater},
  volume = {6},
  number = {1},
  pages = {12},
  publisher = {Nature Publishing Group},
  issn = {2057-3960},
  doi = {10.1038/s41524-020-0276-y},
}

@article{Lee2021,
  title = {Optical Cloaking and Invisibility: {{From}} Fiction toward a Technological Reality},
  shorttitle = {Optical Cloaking and Invisibility},
  author = {Lee, Kyu-Tae and Ji, Chengang and Iizuka, Hideo and Banerjee, Debasish},
  date = {2021-06-15},
  journaltitle = {Journal of Applied Physics},
  shortjournal = {J. Appl. Phys.},
  volume = {129},
  number = {23},
  pages = {231101},
  issn = {0021-8979},
  doi = {10.1063/5.0048846},
}

@online{lee2023,
  title = {{{GELU Activation Function}} in {{Deep Learning}}: {{A Comprehensive Mathematical Analysis}} and {{Performance}}},
  shorttitle = {{{GELU Activation Function}} in {{Deep Learning}}},
  author = {Lee, Minhyeok},
  date = {2023-08-01},
  eprint = {2305.12073},
  eprinttype = {arXiv},
  eprintclass = {cs},
  doi = {10.48550/arXiv.2305.12073},
}

@article{liuGenerativeModelInverse2018,
  title = {Generative {{Model}} for the {{Inverse Design}} of {{Metasurfaces}}},
  author = {Liu, Zhaocheng and Zhu, Dayu and Rodrigues, Sean P. and Lee, Kyu-Tae and Cai, Wenshan},
  date = {2018-10-10},
  journaltitle = {Nano Letters},
  shortjournal = {Nano Lett.},
  volume = {18},
  number = {10},
  pages = {6570--6576},
  publisher = {American Chemical Society},
  issn = {1530-6984},
  doi = {10.1021/acs.nanolett.8b03171},
}

@article{liuTrainingDeepNeural2018,
  title = {Training {{Deep Neural Networks}} for the {{Inverse Design}} of {{Nanophotonic Structures}}},
  author = {Liu, Dianjing and Tan, Yixuan and Khoram, Erfan and Yu, Zongfu},
  date = {2018-04-18},
  journaltitle = {ACS Photonics},
  shortjournal = {ACS Photonics},
  volume = {5},
  number = {4},
  pages = {1365--1369},
  publisher = {American Chemical Society},
  doi = {10.1021/acsphotonics.7b01377},
}

@article{maDeepLearningEnabledOnDemandDesign2018,
  title = {Deep-{{Learning-Enabled On-Demand Design}} of {{Chiral Metamaterials}}},
  author = {Ma, Wei and Cheng, Feng and Liu, Yongmin},
  date = {2018-06-26},
  journaltitle = {ACS Nano},
  shortjournal = {ACS Nano},
  volume = {12},
  number = {6},
  pages = {6326--6334},
  publisher = {American Chemical Society},
  issn = {1936-0851},
  doi = {10.1021/acsnano.8b03569},
}

@article{maProbabilisticRepresentationInverse2019,
  title = {Probabilistic {{Representation}} and {{Inverse Design}} of {{Metamaterials Based}} on a {{Deep Generative Model}} with {{Semi-Supervised Learning Strategy}}},
  author = {Ma, Wei and Cheng, Feng and Xu, Yihao and Wen, Qinlong and Liu, Yongmin},
  date = {2019},
  journaltitle = {Advanced Materials},
  volume = {31},
  number = {35},
  pages = {1901111},
  issn = {1521-4095},
  doi = {10.1002/adma.201901111},
}

@inproceedings{maurya2023,
  title = {Enhancing {{Deep Neural Network Convergence}} and {{Performance}}: {{A Hybrid Activation Function Approach}} by {{Combining ReLU}} and {{ELU Activation Function}}},
  shorttitle = {Enhancing {{Deep Neural Network Convergence}} and {{Performance}}},
  booktitle = {2023 {{Second International Conference}} on {{Informatics}} ({{ICI}})},
  author = {Maurya, Ritesh and Aggarwal, Divyam and Gopalakrishnan, T. and Pandey, Nageshwar Nath},
  date = {2023-11},
  pages = {1--5},
  doi = {10.1109/ICI60088.2023.10421353},
}

@article{pendry2006,
  title = {Controlling {{Electromagnetic Fields}}},
  author = {Pendry, J. B. and Schurig, D. and Smith, D. R.},
  date = {2006-06-23},
  journaltitle = {Science},
  volume = {312},
  number = {5781},
  pages = {1780--1782},
  publisher = {American Association for the Advancement of Science},
  doi = {10.1126/science.1125907},
}

@inproceedings{song2024,
  title = {Comparative {{Analysis}} of {{Activation Functions}} in {{Simple Convolutional Neural Networks}}},
  booktitle = {2024 {{IEEE}} 6th {{International Conference}} on {{Civil Aviation Safety}} and {{Information Technology}} ({{ICCASIT}})},
  author = {Song, Tailin},
  date = {2024-10},
  pages = {1031--1036},
  doi = {10.1109/ICCASIT62299.2024.10827900},
}

@article{Wu2022,
  title = {A Physics-Constrained Deep Learning Based Approach for Acoustic Inverse Scattering Problems},
  author = {Wu, Rih-Teng and Jokar, Mehdi and Jahanshahi, Mohammad R. and Semperlotti, Fabio},
  date = {2022},
  journaltitle = {Mechanical Systems and Signal Processing},
  volume = {164},
  pages = {108190},
  issn = {0888-3270},
  doi = {10.1016/j.ymssp.2021.108190},
}

@article{barnett_evaluation_2014,
	title = {Evaluation of {Layer} {Potentials} {Close} to the {Boundary} for {Laplace} and {Helmholtz} {Problems} on {Analytic} {Planar} {Domains}},
	volume = {36},
	issn = {1064-8275},
	url = {https://doi.org/10.1137/120900253},
	number = {2},
	urldate = {2025-07-02},
	journal = {SIAM J. Sci. Comput.},
	author = {Barnett, Alex H.},
	month = jan,
	year = {2014},
	note = {Publisher: Society for Industrial and Applied Mathematics},
	pages = {A427--A451}
}

@article{Bonnet1999-bf,
  title = {Boundary {{Integral Equation Methods}} for {{Solids}} and {{Fluids}}},
  author = {Bonnet, Marc},
  date = {1999-10-01},
  journaltitle = {Meccanica},
  shortjournal = {Meccanica},
  volume = {34},
  number = {4},
  pages = {301--302},
  issn = {1572-9648},
  doi = {10.1023/A:1004795120236},
}

@article{kress_boundary_1991,
	title = {Boundary integral equations in time-harmonic acoustic scattering},
	volume = {15},
	issn = {0895-7177},
	url = {https://doi.org/10.1016/0895-7177(91)90068-I},
	number = {3},
	journal = {Mathematical and Computer Modelling},
	author = {Kress, Rainer},
	month = jan,
	year = {1991},
	pages = {229--243},
}

@misc{cassier2022activeexteriorcloaking2d,
      title={Active exterior cloaking for the 2D Helmholtz equation with complex wavenumbers and application to thermal cloaking}, 
      author={Maxence Cassier and Trent DeGiovanni and Sébastien Guenneau and Fernando Guevara Vasquez},
      year={2022},
      eprint={2203.02075},
      archivePrefix={arXiv},
      primaryClass={math.AP},
      url={https://arxiv.org/abs/2203.02075}, 
}

@article{metamaterialdesignMachineIntelligence,
  title = {Machine Intelligence in Metamaterials Design: A Review},
  author = {Cerniauskas, Gabrielis and Sadia, Haleema and Alam, Parvez},
  year = 2024,
  month = feb,
  journal = {Oxford Open Materials Science},
  volume = {4},
  number = {1},
  eprint = {https://academic.oup.com/ooms/article-pdf/4/1/itae001/56750204/itae001.pdf},
  pages = {itae001},
  issn = {2633-6979},
  doi = {10.1093/oxfmat/itae001}
}

@article{zhangDataDrivenApproaches2025,
  title = {Data Driven Approaches in Nanophotonics: A Review of {{AI-enabled}} Metadevices},
  shorttitle = {Data Driven Approaches in Nanophotonics},
  author = {Zhang, Huanshu and Kang, Lei and D.~Campbell, Sawyer and T.~Young, Jacob and H.~Werner, Douglas},
  year = 2025,
  journal = {Nanoscale},
  volume = {17},
  number = {41},
  pages = {23788--23803},
  publisher = {Royal Society of Chemistry},
  doi = {10.1039/D5NR02043C},
}

@incollection{InvisibilityCloakingCoordinate2009,
  title = {Invisibility {{Cloaking}} by {{Coordinate Transformation}}},
  author = {Yan, Min and Yan, Wei and Qiu, Min},
  booktitle = {Progress in {{Optics}}},
  year = 2009,
  month = jan,
  volume = {52},
  pages = {261--304},
  publisher = {Elsevier},
  issn = {0079-6638},
  doi = {10.1016/S0079-6638(08)00006-1},
}

@inproceedings{cortesFastAccurateBoundary2024,
  title = {Fast and Accurate Boundary Integral Equation Methods for the Multi-Layer Transmission Problem},
  booktitle = {{{WAVES}} 2024 - {{The}} 16th {{International Conference}} on {{Mathematical}} and {{Numerical Aspects}} of {{Wave Propagation}}},
  author = {Cortes, Elsie A and Carvalho, Camille and Chaillat, St{\'e}phanie and Tsogka, Chrysoula},
  year = 2024,
  month = jun,
  publisher = {Edmond},
  address = {Berlin, Germany},
  doi = {10.17617/3.MBE4AA},
}

@incollection{colton2015inverse,
  title={Inverse scattering},
  author={Colton, D and Kress, R},
  booktitle={Handbook of Mathematical Methods in Imaging: Volume 1, Second Edition},
  pages={649--700},
  year={2015}
}

@article{schweigerFEM1997,
  title = {The Finite-Element Method for the Propagation of Light in Scattering Media: {{Frequency}} Domain Case},
  shorttitle = {The Finite-Element Method for the Propagation of Light in Scattering Media},
  author = {Schweiger, M. and Arridge, S. R.},
  year = 1997,
  journal = {Medical Physics},
  volume = {24},
  number = {6},
  pages = {895--902},
  issn = {2473-4209},
  doi = {10.1118/1.598008},
}

@article{VolakisFEM,
author = {J. L. Volakis and A. Chatterjee and L. C. Kempel},
journal = {J. Opt. Soc. Am. A},
number = {4},
pages = {1422--1433},
publisher = {Optica Publishing Group},
title = {Review of the finite-element method for three-dimensional electromagnetic scattering},
volume = {11},
month = {Apr},
year = {1994},
doi = {10.1364/JOSAA.11.001422},
}

@article{frehnerComparisonFiniteDifference2008,
  title = {Comparison of Finite Difference and Finite Element Methods for Simulating Two-Dimensional Scattering of Elastic Waves},
  author = {Frehner, Marcel and Schmalholz, Stefan M. and Saenger, Erik H. and Steeb, Holger},
  year = 2008,
  month = dec,
  journal = {Physics of the Earth and Planetary Interiors},
  series = {Recent {{Advances}} in {{Computational Geodynamics}}: {{Theory}}, {{Numerics}} and {{Applications}}},
  volume = {171},
  number = {1},
  pages = {112--121},
  issn = {0031-9201},
  doi = {10.1016/j.pepi.2008.07.003},

}

@article{ra2001problems,
  title={Problems in Inverse Scattering-Illposedness, Resolution, Local Minima, and Uniquenesse},
  author={Ra, Jung-Woong},
  journal={Communications of the Korean Mathematical Society},
  volume={16},
  number={3},
  pages={445--458},
  year={2001},
  publisher={Korean Mathematical Society}
}

@article{liuBEMAcousticWave2019,
  title = {On the {{BEM}} for Acoustic Wave Problems},
  author = {Liu, Yijun},
  year = 2019,
  month = oct,
  journal = {Engineering Analysis with Boundary Elements},
  volume = {107},
  pages = {53--62},
  issn = {0955-7997},
  doi = {10.1016/j.enganabound.2019.07.002},
  urldate = {2025-07-21},
}

@misc{CCCT2026,
  title = {Discretization in {{Multilayered Media}} with {{High Contrasts}}: {{Is It All About}} the {{Boundaries}}?},
  shorttitle = {Discretization in {{Multilayered Media}} with {{High Contrasts}}},
  author = {Carvalho, Camille and Chaillat, St{\'e}phanie and Cortes, Elsie and Tsogka, Chrysoula},
  year = 2026,
  month = feb,
  number = {arXiv:2602.18629},
  eprint = {2602.18629},
  primaryclass = {math},
  publisher = {arXiv},
  }

@article{sunBoundaryRegularizedIntegral2015,
  title = {Boundary Regularized Integral Equation Formulation of the {{Helmholtz}} Equation in Acoustics},
  author = {Sun, Qiang and Klaseboer, Evert and Khoo, Boo-Cheong and Chan, Derek Y. C.},
  year = 2015,
  month = jan,
  journal = {Royal Society Open Science},
  volume = {2},
  number = {1},
  pages = {140520},
  issn = {2054-5703},
}

@article{carvalhoAsymptotic2020,
author = {Carvalho, Camille and Khatri, Shilpa and Kim, Arnold D.},
title = {Asymptotic Approximations for the Close Evaluation of Double-Layer Potentials},
journal = {SIAM Journal on Scientific Computing},
volume = {42},
number = {1},
pages = {A504-A533},
year = {2020},
doi = {10.1137/18M1218698},
}

@article{perezPlanewave2019,
author = {P\'{e}rez-Arancibia, Carlos and Turc, Catalin and Faria, Luiz},
title = {Planewave Density Interpolation Methods for 3D Helmholtz Boundary Integral Equations},
journal = {SIAM Journal on Scientific Computing},
volume = {41},
number = {4},
pages = {A2088-A2116},
year = {2019},
doi = {10.1137/19M1239866},
}

@article{klocknerQuadratureExpansionNew2013,
  title = {Quadrature by Expansion: {{A}} New Method for the Evaluation of Layer Potentials},
  shorttitle = {Quadrature by Expansion},
  author = {Kl{\"o}ckner, Andreas and Barnett, Alexander and Greengard, Leslie and O'Neil, Michael},
  year = 2013,
  month = nov,
  journal = {Journal of Computational Physics},
  volume = {252},
  pages = {332--349},
  issn = {0021-9991},
  doi = {10.1016/j.jcp.2013.06.027},

}

\end{document}